\documentclass[a4paper,epj,nopacs]{svjour}
\usepackage[reqno]{amsmath}
\usepackage{amsfonts,amssymb,latexsym}
\usepackage{graphicx}

\newcommand{\psibar}{\bar{\psi}}
\newcommand{\half}{\frac{1}{2}}
\newcommand{\bra}{\langle}
\newcommand{\ket}{\rangle}
\newcommand{\tr}{\operatorname{Tr}}
\newcommand{\order}{{\cal O}}
\newcommand{\One}{1\kern-4.5pt1}
\newcommand{\that}{\hat{t}}
\newcommand{\kt}{\tilde{k}}
\newcommand{\Udag}{U^\dagger}
\newcommand{\eq}{\varepsilon_q}
\newcommand{\eg}{\varepsilon_g}
\newcommand{\qq}{\bra qq\ket}

\newcommand{\latt}{{\text{lat}}}
\newcommand{\cont}{{\text{cont}}}
\newcommand{\pbp}{\bra\psibar\psi\ket}

\newcommand{\cadj}{\cite{Hands:2000ei,Hands:2001ee}}
\newcommand{\cwilsa}{\cite{Skullerud:2003yc}}
\newcommand{\ckstvz}{\cite{Kogut:2000ek}}
\newcommand{\ckshm}{\cite{Kogut:2001na}}

\newcommand{\ccscrev}{\cite{Rajagopal:2000wf,Alford:2001dt,Schafer:2003vz,Shovkovy:2004me}}
\newcommand{\chight}{\cite{Philipsen:2005mj}}
\newcommand{\cltcqcd}{\cite{Hands:1999md,Aloisio:2000if,Hands:2000ei,Hands:2001ee,Kogut:2002cm,Muroya:2002ry,Alles:2006ea,Chandrasekharan:2006tz}}
\newlength{\colw}
\begin{document}
\bibliographystyle{h-physrev4}

\title{Deconfinement in dense 2-color QCD}

\author{Simon Hands\inst{1} \and Seyong Kim\inst{2} \and Jon-Ivar
  Skullerud\inst{3}}

\institute{Department of Physics, University of Wales Swansea,
 Singleton Park, Swansea SA2 8PP, UK
\and Department of Physics, Sejong University, Gunja-Dong, Gwangjin-Gu,
Seoul 143-747, Korea
\and School of Mathematics, Trinity College, Dublin 2, Ireland}

\abstract{
We study SU(2) lattice gauge theory with two flavors of Wilson
fermion at non-zero chemical potential $\mu$ and low temperature on a
$8^3\times16$ system.  We
identify three r\'egimes along the $\mu$-axis.  For $\mu\lesssim
m_\pi/2$ the system remains in the vacuum phase, and all physical
observables considered remain essentially unchanged.  The intermediate
r\'egime is characterised by a non-zero diquark condensate and an
associated increase in the baryon density, consistent with what is
expected for Bose--Einstein condensation of tightly bound diquarks.
We also observe screening of the static quark potential here.  In the
high-density deconfined r\'egime we find a non-zero Polyakov loop and
a strong modification of the gluon propagator, including significant screening
in the magnetic sector in the static limit, which must have a non-perturbative
origin. The behaviour of thermodynamic observables and the superfluid order
parameter are consistent
with a Fermi surface disrupted by a BCS diquark condensate.
The energy per baryon as a function of $\mu$ exhibits a minimum 
in the deconfined r\'egime, implying
that macroscopic objects such as stars formed in this theory are largely
composed of quark matter.
} \maketitle

\section{Introduction}

At large quark chemical potential $\mu$, QCD is expected to undergo a
transition from a confined nuclear matter phase to a deconfined quark
matter phase, where the relevant degrees of freedom are quarks and
gluons.  It is now also generally believed that the quark matter phase
at low temperature $T$ is characterised by diquark condensation:
pairing of quarks near the Fermi surface gives rise to a number of
color superconducting phases \ccscrev.  The phase structure
depends critically on the precise values of the diquark gap parameters
and the effective strange quark mass at the relevant densities, and in
the absence of a first-principles non-perturbative determination of
these quantities, our knowledge of this region of the phase diagram
will remain unsatisfactory.

Lattice QCD is at present unable to address these problems directly,
since the fermion determinant is no longer positive definite once
$\mu\neq0$, and cannot be used as a probability weight in the
functional integral.  There has been much progress in recent years in
developing methods for the region of high $T$, low $\mu$, where the
problem is less severe \chight.  These methods are being extended
in the direction of higher $\mu$ and lower $T$, but it is not clear
at what point they break down, beyond which only unphysical results will
be obtained.

Another approach is to study QCD-like theories where the fermion
determinant remains real and positive even when $\mu\neq0$.  These can be
used as a laboratory for investigating gauge theories at high density
and low temperature.  One such theory is 2-color QCD (QC$_2$D) with an even
number of flavors \cite{Hands:1999md}, where the quarks and antiquarks live
in equivalent representations of the color group and can be related by
an anti-unitary symmetry (the Pauli--G\"ursey symmetry).  This theory
has been studied on the lattice by a number of groups \cltcqcd.

At $\mu=0$, the Pauli--G\"ursey symmetry implies an exact symmetry
between mesons and diquarks, which are the baryons of the theory.  In
particular, chiral multiplets will contain both mesons and baryons.
For $N_f=2$ for example, the pseudo-Goldstone multiplet consists of
the pion isotriplet plus a scalar isoscalar diquark and antidiquark.
The diquark baryons can be expected to condense when $\mu\gtrsim
m_\pi/2$, forming a superfluid ground state.  In this respect, the
theory is radically different from real QCD, where no gauge invariant
diquark operator exists and the ground state at high density is
superconducting.  The nature of the superfluid ground state is however
an interesting issue in its own right.  For instance, an alternative
approach to a superfluid order parameter in terms of an orthodox BCS
description of diquark pairing at the Fermi surface has been given in
\cite{Kondratyuk:1991hf}.

In the gluon sector, the differences between SU(2) and SU(3) are
expected to be less important, and QC$_2$D is a good setting for
{\em ab initio} studies of gluodynamics in the presence of a background
baryon density.  Of particular interest is the issue of deconfinement
at high density.  Signals of deconfinement were observed in
simulations with Wilson \cite{Muroya:2002ry} and staggered fermions
\cite{Alles:2006ea}, where correlations were found between the
Polyakov loop and chiral or baryonic observables.  However, the phase
structure has not been investigated in any further detail, and it
remains unclear whether there is a confined nuclear matter phase 
with non-zero baryon
number (as in QCD), or just a single phase transition.  This will be
investigated in the present paper.

Most of the lattice studies so far have been conducted using staggered
fermions, which have several potential theoretical problems.  Firstly,
the pattern of global symmetry breaking is different to that of the
continuum, so two-color staggered lattice QCD has a different
Goldstone spectrum to continuum QC$_2$D \cadj.  Secondly, one
staggered flavor corresponds to four continuum flavors, which is
uncomfortably close to the Banks--Zaks threshold
$N_f^{BZ}=136/49\approx5.6$ where the second term in the
$\beta$-function changes sign, leading to a nontrivial infrared fixed
point and absence of confinement and chiral symmetry breaking.  Note that 
this prediction for $N_f^{BZ}$ is inherently perturbative, and may be unreliable
since the fixed-point coupling is large for $N_f\sim N_f^{BZ}$.
In
order to describe a single continuum flavor one usually takes the
fourth root of the fermion determinant.  It is not clear whether this
procedure yields something corresponding to a local action of a
single-flavor fermion field \cite{Durr:2005ax}.  Even if it can be
shown to be a valid procedure at $\mu=0$, obstacles remain for
$\mu\neq0$ which may invalidate it \cite{Golterman:2006rw}.

These problems are absent if Wilson fermions are used. We note that the Wilson
formulation still admits a U(1)$_B$ global symmetry implying a 
conserved baryon number, and so a superfluid order parameter remains
well-defined. In the chiral limit $\kappa\to\kappa_c$, 
the lattice Dirac operator's eigenvalue spectrum lies in the same 
chiral orthogonal ensemble universality class as the continuum theory. 
Moreover, the consequences of explicitly broken chiral
symmetry should be less severe, since they manifest themselves at the bottom of
the Fermi sea, and hence become physically irrelevant at large $\mu$.
Because the Wilson formulation has the whole of the first 
Brillouin zone available to describe a single physical flavor, saturation
artefacts set in at a larger value of $\mu$ than is the case for
staggered \cite{Bietenholz:1998rj}.
On the other
hand one has to contend with a higher computational cost,  
and for this reason, only
a few studies using Wilson fermions have been performed to date
\cite{Muroya:2002ry}.

In this article we will present results from a study of QC$_2$D
with two flavors of Wilson fermion at zero temperature and non-zero
chemical potential.  In section~\ref{sec:formulation} we present the
lattice action and expressions for the principal bulk observables.  The
simulation parameters are given in section~\ref{sec:params}, along
with results for the lattice spacing and pion and rho meson masses
from simulations at $\mu=0$.  The main results are given in
section~\ref{sec:results}.  Finally, in section~\ref{sec:discussion}
we discuss the significance of our results and the prospects for
further work.

\section{Lattice formulation and simulation}
\label{sec:formulation}

\subsection{Action and algorithm}

The $N_f=2$ fermion action is given by
\begin{multline}
S=\bar\psi_1M(\mu)\psi_1+\bar\psi_2M(\mu)\psi_2\\
-J\bar\psi_1(C\gamma_5)\tau_2\bar\psi_2^{tr}
+\bar J\psi_2^{tr}(C\gamma_5)\tau_2\psi_1,\label{eq:action}
\end{multline}
where $M(\mu)$ is the usual Wilson fermion matrix
\begin{multline}
M_{xy}(\mu)=\delta_{xy}
-\kappa\sum_{\nu}\Bigl[(1-\gamma_\nu)e^{\mu\delta_{\nu0}}U_\nu(x)\delta_{y,x+\hat\nu}\\
+(1+\gamma_\nu)e^{-\mu\delta_{\nu0}}U_\nu^\dagger(y)\delta_{y,x-\hat\nu}\Bigr]\,.
\end{multline}
The diquark source terms $J,\bar{J}$ serve a double purpose in lifting
the low-lying eigenmodes in the superfluid phase, thus making the
simulation numerically tractable, and enabling us to study diquark
condensation without any ``partial quenching''.  In principle the results 
should at
the end be extrapolated to the ``physical'' limit $J=\bar{J}=0$.  We will
also introduce the rescaled source strength $j\equiv J/\kappa$.

The fermion matrix has the following symmetries:
\begin{align}
\gamma_5M(\mu)\gamma_5&=M^\dagger(-\mu)\label{eq:dagger}\\
KM(\mu)K^{-1} &= M^*(\mu)\quad\text{with}\quad K\equiv
C\gamma_5\tau_2\,,
\label{eq:PG}
\end{align}
where we have used the property $\tau_2 U_\mu\tau_2=U_\mu^*$.

The last equation is the Pauli--G\"ursey symmetry.
This symmetry implies that $\det M(\mu)$ is real, but not necessarily
positive.  However, with the change of variables $\bar\phi=-\psi_2^{tr}C\tau_2,
\phi=C^{-1}\tau_2\bar\psi_2^{tr}$ we can rewrite the action as
\begin{equation}
S=\begin{pmatrix}\bar\psi &\bar\phi\end{pmatrix}
\begin{pmatrix} M(\mu)&J\gamma_5\\
            -\bar J\gamma_5&M(-\mu)\end{pmatrix}
\begin{pmatrix}\psi\\\phi\end{pmatrix}\equiv\bar\Psi{\cal M}\Psi.
\label{eq:bilinear}
\end{equation}
Hence positivity of $\mbox{det}{\cal M}$ requires the product $J\bar J$ to be
real and positive, which translates into the requirement that the diquark
source term be antihermitian \cwilsa.

Now use (\ref{eq:bilinear}) to write
\begin{equation}
{\cal M}^\dagger{\cal M}=
\begin{pmatrix}M^\dagger(\mu)M(\mu)\!+\!|\bar J|^2&\\
&\!\!\!\!\!M^\dagger(-\mu)M(-\mu)\!+\!|J|^2
\end{pmatrix}
\end{equation}
The off-diagonal terms can be shown to vanish if $\bar J=J^*$
using (\ref{eq:dagger}); moreover
the same identity applied to the lower block yields
\begin{equation}
\det{\cal M}^\dagger{\cal M}=[\det(M^\dagger(\mu)M(\mu)+\bar{J}J)]^2.
\end{equation}
It is therefore possible to take the square root analytically, by
using pseudofermion fields with weight $(M^\dagger M+|J|^2)^{-1}$.
This has the advantage of {\em(a)} requiring matrix multiplications of
half the dimensionality, and {\em(b)} permitting a hamiltonian
evaluation and hence the use of an exact HMC algorithm.  It is
equivalent to the even/odd partitioning step used for staggered
fermion gauge theories, but is more transparent since all lattice
sites are physically equivalent, making the force term easier to
implement.  The trick was used in \ckshm, though because the staggered
version still requires a Pfaffian rather than a determinant, an HMD
algorithm was used in that case.

\subsection{Observables}

The quark number density is given by the timelike component of the
conserved vector current:
\begin{multline}
n_q =
\sum_i\kappa\Bigl\bra\psibar_i(x)(\gamma_0-1)e^{\mu}U_t(x)\psi_i(x+\that)\\
+ \psibar_i(x)(\gamma_0+1)e^{-\mu}\Udag_t(x-\that)\psi_i(x-\that)\Bigr\ket\,.
\label{eq:nq}
\end{multline}
The quark energy density can be defined in terms of a local bilinear
very similar to (\ref{eq:nq}):
\begin{multline}
\eq =
\sum_i\kappa\Bigl\bra\psibar_i(x)(\gamma_0-1)e^{\mu}U_t(x)\psi_i(x+\that)\\
- \psibar_i(x)(\gamma_0+1)e^{-\mu}\Udag_t(x-\that)\psi_i(x-\that)\Bigr\ket\,.
\label{eq:eq}
\end{multline}
Unlike $n_q$, however, this quantity requires both additive and multiplicative
renormalisation as a result of quantum corrections. First, the vacuum
contribution $\varepsilon_{q0}$ must be subtracted. This correction can be
obtained in the zero temperature thermodynamic limit from the relation
\begin{equation}
\eq^0 = \frac{1}{D}\Bigl(\tr\One-\bra\psibar\psi\ket_{\mu=0}\Bigr)\,,
\label{eq:eq-subtract}
\end{equation}
valid in $D$-dimensional spacetime, or more directly, as we do here,
by subtracting the measured value of $\eq(\mu=0)\simeq0.3982(8)$.  We
have verified that this gives the same result as using
eq.~(\ref{eq:eq-subtract}).  The multiplicative correction results
from the renormalisation of the anisotropy factor $\xi=a_t/a_s$ under
quantum corrections; it affects the quantity defined in
(\ref{eq:eq}) even in the isotropic limit $\xi=1$, and must be
determined either perturbatively or preferably non-perturbatively,
requiring simulations with $\xi\neq1$. To our knowledge the
perturbative correction has yet to be calculated for Wilson fermions;
for staggered fermions it has been computed to be
$1-0.1599C_2(N_c)g^2$ \cite{Karsch:1989fu}, where $C_2$ is the
quadratic Casimir. This suggests that our results for $\eq$ should be
rescaled downwards. Since the correction is independent of $\mu$,
however, this is only an overall normalisation factor.

With the standard Wilson lattice gauge action employed in this study
the gluon energy density may be
defined as a local observable
\begin{equation}
\varepsilon_g\equiv
\frac{1}{N_s^3N_t}
\left\bra a_t\frac{\partial S_g}{\partial a_t}\right\ket
=\frac{3\beta}{N_c}\tr\bra \Box_t-\Box_s\ket,
\end{equation}
where $\Box_t$, $\Box_s$ are timelike and spacelike plaquettes respectively.
Once again, this requires renormalisation due to quantum corrections;
the dominant correction factor has been calculated in perturbation
theory to be $1-0.1466C_2g^2$ \cite{Karsch:1982ve}, suggesting that
bare $\eg$ data should again be rescaled downwards by a
$\mu$-independent factor.

The final thermodynamic observable we can discuss is the trace of the
energy-momentum tensor, expressible in terms of the conformal anomaly
\begin{equation}
\begin{split}
\delta&=\varepsilon-3p\\
&=-\frac{1}{N_s^3N_t}\Biggl[
a\frac{\partial\beta}{\partial a}\biggr\vert_{LCP}
\!\!\!\frac{\partial\ln{\cal Z}}{\partial\beta}
+a\frac{\partial\kappa}{\partial a}\biggr\vert_{LCP}
\!\!\!\frac{\partial\ln{\cal Z}}{\partial\kappa} \\
& \phantom{=-\frac{1}{N_s^3N_t}\Biggl[}
 +a\frac{\partial j}{\partial a}\biggr\vert_{LCP}
\!\!\!\frac{\partial\ln{\cal Z}}{\partial j}
\Biggr]
\end{split}
\label{eq:conformal}
\end{equation}
where it is understood that beta-functions are evaluated at $\mu=T=0$
along lines of constant physics, so dimensionless ratios of physical
quantities are cutoff-independent (the derivation of this equation for
$\mu\neq0$ is sketched in \cite{Allton:2003vx}). Baryon number
symmetry implies that $\lim_{j\to0}\partial j/\partial a=0$ and hence
the third term can be neglected.  The local observables required at
$\mu\neq0$ are then
\begin{align}
-\frac{1}{N_s^3N_t}\frac{\partial\ln{\cal Z}}{\partial\beta} &=
-\frac{3\beta}{N_c}\tr\bra \Box_t+\Box_s\ket \label{eq:conformg}\\
-\frac{1}{N_s^3N_t}\frac{\partial\ln{\cal Z}}{\partial\kappa} &=
\frac{1}{\kappa}(\tr\One-\bra\bar\psi\psi\ket) \label{eq:conformq}
\end{align}

The final bulk observables we compute are the diquark condensate,
\begin{equation}
\bra qq\ket = \frac{\kappa}{2}\bra\psibar_1K\psibar_2^{tr} -
\psi_2^{tr}K\psi_1\ket\,,
\label{eq:qq}
\end{equation}
which is an order parameter for the vacuum-to-superfluid transition;
and the Polyakov loop $N_c^{-1}\bra\tr L\ket$, which in pure gauge theories
is an order parameter for
the deconfinement transition.

\section{Simulation parameters and physical scales}
\label{sec:params}

Exploratory simulations at zero chemical potential were performed at a
range of different values for $\beta$ and $\kappa$ \cwilsa, but for
this study the parameters $\beta=1.7, \kappa=0.178$ were selected.
All the simulations have been performed on a $8^3\times16$ lattice.
The full set of parameters used is given in table~\ref{tab:params}.
Configurations were saved every 4 trajectories.
\begin{table}
\begin{tabular*}{\colw}{l@{\extracolsep{\fill}}llllrl}
\hline
$\mu$ & $j$ & $N_{traj}$ & $\bra\ell\ket$ & $dt$ & $N_{cg}$ & acc \\ \hline
0    & 0   & 2000 & 1.0 & 0.0125 &  385 & 85\% \\
0.1  & 0    & 608 & 1.0 & 0.0125 &      & 82\% \\
0.2  & 0    & 560 & 1.0 & 0.01   &  785 & 75\% \\
0.25 & 0.04 & 266 & 0.5 & 0.0075 &  740 & 86\% \\
0.3  & 0    & 600 & 1.0 & 0.004  & 1350 & 80\% \\
     & 0.02 & 188 & 0.5 & 0.006  & 1170 & 80\% \\
     & 0.04 & 190 & 0.5 & 0.0075 &  965 & 75\% \\
     & 0.06 & 276 & 0.5 & 0.0075 &  775 & 86\% \\
0.35 & 0    & 500 & 0.1 & 0.0005 & 1650 & 90\% \\
     & 0.02 & 400 & 1.0 & 0.004  & 1615 & 75\% \\
     & 0.04 & 500 & 1.0 & 0.005  & 1090 & 85\% \\
0.4  & 0.04 & 148 & 1.0 & 0.005  & 1165 & 80\% \\
     &      & 302 & 0.5 & 0.006  & 1235 & 77\% \\
0.45 & 0.04 & 252 & 0.5 & 0.0042 & 1275 & 82\% \\
     &      &  64 & 1.0 & 0.004  & 1200 & 88\% \\
0.5  & 0.02 & 554 & 0.5 & 0.003  & 2565 & 78\% \\
     & 0.04 &  44 & 1.0 & 0.004  & 1240 & 74\% \\
     &      & 300 & 0.5 & 0.0045 & 1270 & 77\% \\
     & 0.06 & 304 & 0.5 & 0.005  &  900 & 86\% \\
0.55 & 0.04 & 308 & 0.5 & 0.0038 & 1340 & 83\% \\
0.6  & 0.04 & 112 & 1.0 & 0.0033 & 1290 & 83\% \\
     &      & 283 & 0.5 & 0.004  & 1375 & 80\% \\
0.65 & 0.04 & 276 & 1.0 & 0.003  & 1310 & 85\% \\
0.7  & 0.02 & 400 & 1.0 & 0.002  & 2525 & 80\% \\
     &      & 440 & 0.5 & 0.0025 & 2760 & 75\% \\
     & 0.04 & 136 & 1.0 & 0.003  & 1330 & 85\% \\
     &      & 530 & 0.5 & 0.0035 & 1450 & 78\% \\
     & 0.06 & 280 & 0.5 & 0.0037 &  970 & 85\% \\
0.75 & 0.04 & 180 & 1.0 & 0.003  & 1390 & 80\% \\
0.8  & 0.04 & 292 & 1.0 & 0.0028 & 1410 & 80\% \\
     &      & 292 & 0.5 & 0.003  & 1520 & 74\% \\
0.9  & 0.04 & 264 & 0.5 & 0.003  & 1610 & 72\% \\
1.0  & 0.04 & 272 & 0.5 & 0.0028 & 1660 & 75\% \\ 
1.3  & 0.04 & 249 & 0.5 & 0.0033 & 1070 & 88\% \\
1.4  & 0.04 & 404 & 0.5 & 0.005  & 620  & 82\% \\
1.5  & 0.04 & 400 & 0.5 & 0.01   & 290  & 80\% \\
1.6  & 0.04 & 400 & 0.5 & 0.01   & 215  & 90\% \\
1.75 & 0.04 & 250 & 0.5 & 0.01   & 200  & 94\% \\
2.5  & 0.04 & 260 & 0.5 & 0.012  & 260  & 97\% \\ \hline

\end{tabular*}
\caption{Simulation parameters. $\bra\ell\ket$ denotes the average
  trajectory length, $dt$ is the molecular dynamics timestep, $N_{cg}$
  is the number of conjugate gradient iterations needed in the
  molecular dynamics evolution, and acc is the acceptance rate.}
\label{tab:params}
\end{table}

\begin{figure}
\includegraphics*[width=\colw]{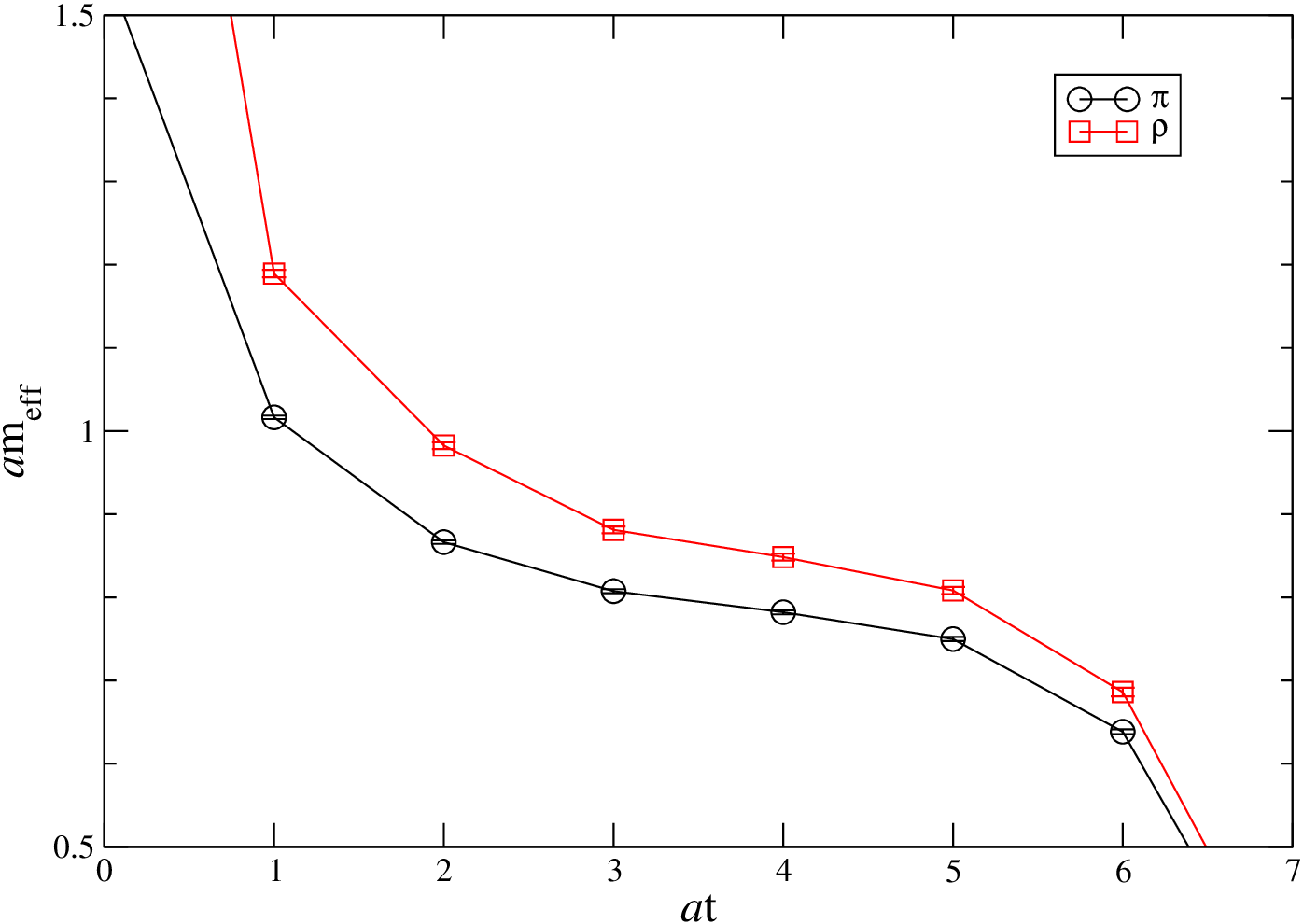}
\caption{Pi and rho effective masses for $\mu=j=0$.}
\label{fig:pirho}
\end{figure}

In order to determine the scale, we computed pi and rho meson
correlators and the static quark potential.  The mesons were computed
using point sources (no smearing), and as fig.~\ref{fig:pirho} and
table~\ref{tab:pirhomass} clearly show, on our relatively small
lattice this did not allow a precise determination of the mass.  In a
future in-depth study of the hadron spectrum it will be desirable to
use smeared sources and variational techniques, but at this point we
are primarily interested in a rough idea of the hadronic scales.
\begin{table}
\begin{tabular}{lllllll}
\hline
Range & $m_\pi$ & $\chi^2/N_{df}$ & $m_\rho$ & $\chi^2/N_{df}$ &
$m_\pi/m_\rho$ \\ \hline
3--4 & 0.809(2) & 0.58 & 0.882(4) & 0.71 & 0.917(3) \\
3--5 & 0.800(2) & 0.97 & 0.870(3) & 1.4 & 0.919(3) \\
3--6 & 0.795(2) & 1.3 & 0.860(3) & 2.5 & 0.924(3) \\
4--5 & 0.789(2) & 0.81 & 0.854(4) & 1.7 & 0.924(3) \\
4--6 & 0.786(2) & 0.63 & 0.846(3) & 1.9 & 0.929(3) \\
5--6 & 0.783(2) & 0.80 & 0.836(4) & 3.0 & 0.937(3) \\ \hline
\end{tabular}
\caption{Fit ranges and fitted masses for pi and rho mesons.}
\label{tab:pirhomass}
\end{table}
These numbers indicate that we may expect an onset transition at
$\mu_o\approx m_\pi/2\sim 0.4a^{-1}$, but since there is no separation
of scales between the Goldstone (pion) and the non-Goldstone (rho), we
should not expect chiral perturbation theory to be quantitatively
valid at any point.

\begin{figure}
\includegraphics*[width=\colw]{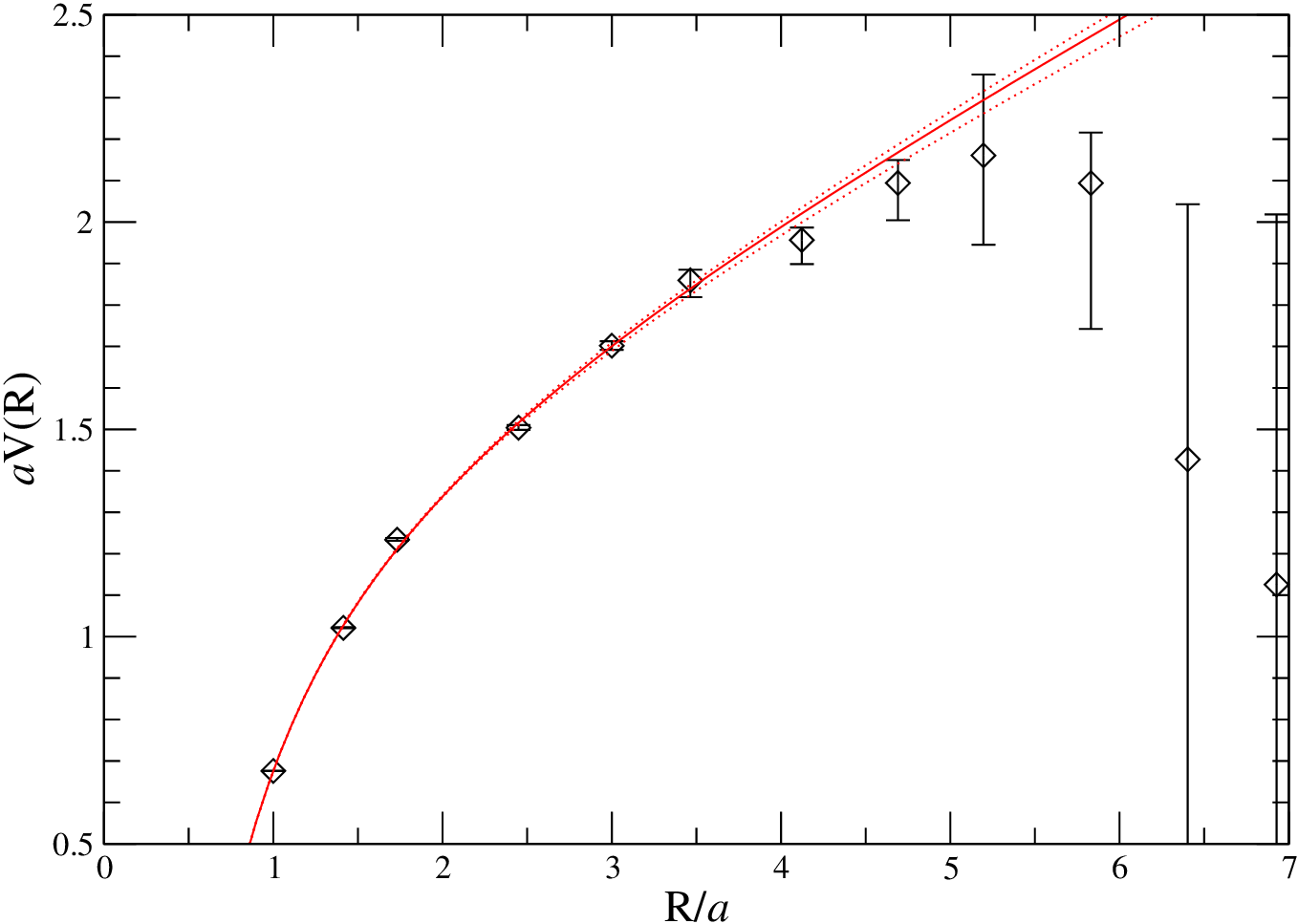}
\caption{Static quark potential for $\mu=j=0$, together with a fit to
  the Cornell potential.}
\label{fig:V0}
\end{figure}

The static quark potential was computed using APE smeared Wilson loops
with spatial separations near the diagonal to minimise lattice
artefacts. The minimum time separation used was 2.
Figure~\ref{fig:V0} shows the potential together with a fit to the
Cornell potential $V(R)=\sigma R+e/R+C$ up to $aR_{\text{max}}=4.0$.  We
find for the string tension $\sigma a^2=0.218(8)$, which gives
$a=0.223(4)$ fm for the lattice spacing.

\section{Results at $\mu\neq0$}
\label{sec:results}

\subsection{Model Considerations}
\label{sec:models}

Our expectation as $\mu$ is increased from zero at $T\simeq0$ is that
the system will remain in the vacuum phase until an onset occurs at
$\mu=\mu_o\simeq M_{\text{light}}/N_c$, where $M_{\text{light}}$ is
the mass of the lightest baryon in the physical spectrum. In 
QC$_2$D this lightest baryon is a scalar diquark state in the same
chiral multiplet as the pion: hence $\mu_o=m_\pi/2$, and in the chiral
limit the onset transition is well described by an effective chiral
model ($\chi$PT) in which only pseudo-Goldstone pion and diquark
degrees of freedom are retained. A mean-field treatment of $\chi$PT
has yielded quantitative predictions for chiral and superfluid
condensates, quark number density, and Goldstone spectrum as $\mu$ is
increased beyond onset \ckstvz.  Our starting point for thermodynamics
is the result for quark number density:
\begin{equation}
n_{\chi PT}=
\begin{cases}
  0\,, & \mu<\mu_o,\\
  8N_fF^2\mu\left(1-\frac{\mu_o^4}{\mu^4}\right)\,, & \mu\geq\mu_o.
\end{cases}
\label{eq:nchiPT}
\end{equation}
Here $F$ is the pion decay constant, a parameter of the $\chi$PT which
can be extracted in principle from the pion correlator measured at
$\mu=0$.  The pressure follows immediately from integration of the
fundamental relation $n_q={\partial p/\partial\mu}\vert_{T,V}$:
\begin{equation}
p_{\chi PT}=
4N_fF^2\left(\mu^2+{\frac{\mu_o^4}{\mu^2}}-2\mu_o^2\right).
\label{eq:pchiPT}
\end{equation}
Now, if $\Omega(\mu,T,V)$ is the thermodynamic grand potential, then in the 
grand canonical ensemble $\Omega=-pV$. In the $T=0$ limit we can therefore
extract the energy density via $\Omega/V=\eq-\mu n_q$,
implying
\begin{equation}
\varepsilon_{\chi PT}=4N_fF^2\left(\mu^2-3\frac{\mu_o^4}{\mu^2}+2\mu_o^2\right).
\label{eq:epschiPT}
\end{equation}
We can also extract the trace of the energy-momentum tensor
$\delta=\eq-3p$:
\begin{equation}
\delta_{\chi PT}=8N_fF^2\left(-\mu^2-3\frac{\mu_o^4}{\mu^2}+4\mu_o^2\right)
\end{equation}
and note that $\delta_{\chi PT}$ is positive for $\mu_o<\mu<\sqrt{3}\mu_o$.

The result (\ref{eq:epschiPT}) for $\varepsilon_{\chi PT}$ coincides
with that derived in \ckstvz\ using the assumption that diquarks
Bose-condense in the ground state, remain degenerate with the pion for
$\mu\geq\mu_o$, and that their self-interactions are weak due to their
Goldstone nature.  One infers $\eq\simeq\half m_\pi n_q=\mu_o n_q$.  A
more refined treatment taking account of corrections to the dilute
ideal Bose gas from the weak repulsive interaction yields \ckstvz
\begin{equation} 
\varepsilon_q=\mu_on_q+{\frac{n_q^2}{64F^2N_f}}+\cdots
\label{eq:DBG}
\end{equation}
where the dots denote corrections from subleading terms in the chiral
expansion.  Using $\mu=\partial\eq/\partial n_q$, this can be seen to
be consistent with (\ref{eq:nchiPT}) linearised about $\mu=\mu_o$:
\begin{equation}
n_{\chi PT}\approx32N_fF^2(\mu-\mu_o);
\end{equation}
the same approximation predicts $p_{\chi PT}$ to vanish as
$(\mu-\mu_o)^2$, consistent with a second order transition in the
Ehrenfest scheme.

Next we turn to the deconfined phase expected at large $\mu$. The obvious
starting point is the Stefan--Boltzmann prediction for the number density
of massless quarks:
\begin{equation}
n_{SB}=\frac{N_fN_c}{3\pi^2}\mu^3.
\label{eq:nSB}
\end{equation}
This is obtained simply by populating a Fermi sphere of radius
$k_F=\mu$, with every momentum state occupied by $2N_fN_c$
non-interacting quarks. Other thermodynamic quantities follow
immediately:
\begin{equation}
\varepsilon_{SB}=3p_{SB}={\frac{N_fN_c}{4\pi^2}}\mu^4;\quad\delta_{SB}=0.
\label{eq:thermoSB}
\end{equation}
An estimate for the chemical potential $\mu_d$ at which deconfinement
takes place can be obtained by equating the free grand potential
densities, or equivalently the pressures given in (\ref{eq:pchiPT})
and (\ref{eq:thermoSB}).  Since $p_{\chi PT}>p_{SB}$ for
$\mu<\mu_d$ is required for thermodynamic stability, we find $\mu_d$
given by the largest positive real root of
\begin{equation}
\mu_d^3-4\pi F\sqrt{\frac{3}{N_c}}(\mu_d^2-\mu_o^2)=0.
\end{equation}
This estimate takes no account of any non-Goldstone states in the hadron
spectrum, or of any gluon degrees of freedom released at deconfinement.
\begin{figure}
\begin{center}
\includegraphics*[width=\colw]{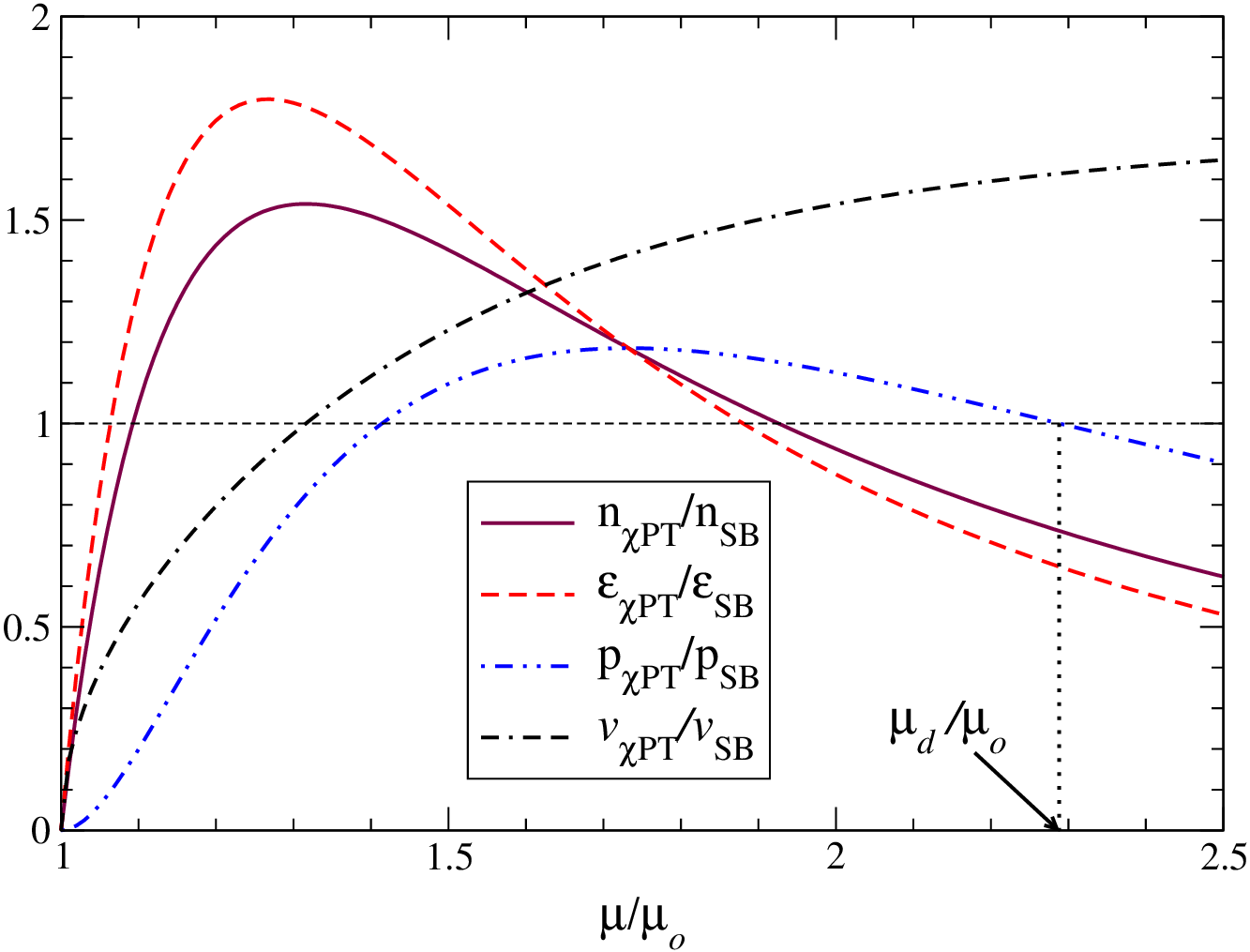}
\end{center}
\caption{Ratio of thermodynamic observables
from $\chi$PT and free quarks}
\label{fig:model}
\end{figure}
In fig.~\ref{fig:model} we plot the ratios $n_{\chi PT}/n_{SB}$,
$\varepsilon_{\chi PT}/\varepsilon_{SB}$ and $p_{\chi PT}/p_{SB}$ as functions
of $\mu/\mu_o$ for the choice $F^2=N_c/6\pi^2$, corresponding to
$\mu_d\simeq2.288\mu_o$. Since $n_{\chi PT}<n_{SB}$ at this point,
this naively simple approach predicts a first
order deconfining transition --- note also that $\delta_{\chi PT}<0$ 
at deconfinement.
Also shown is the ratio of the speed of sound
\begin{equation}
v_{\chi PT}=\sqrt{\frac{\partial p}{\partial\varepsilon}}=
\sqrt{\frac{1-\frac{\mu_o^4}{\mu^4}}{1+3\frac{\mu_o^2}{\mu^4}}}
\end{equation}
to the Stefan--Boltzmann value $v_{SB}=1/\sqrt{3}$.

\subsection{Thermodynamics results}
\label{sec:thermo}

The raw data for bosonic observables (spatial and temporal plaquettes,
Polyakov line) are tabulated in table~\ref{tab:gluon}, and for
fermionic bilinears ($\pbp, n_q$ (\ref{eq:nq}), $\eq$ (\ref{eq:eq}) and $\qq$
(\ref{eq:qq})) in table~\ref{tab:quark}. All quark observables are
normalised to $N_c$ colors and $N_f$ flavors, with $N_f=N_c=2$. In
this section we outline the analysis needed to extract bulk
thermodynamical quantities and condensates.
\begin{table*}
\begin{tabular*}{0.9\textwidth}{l@{\extracolsep{\fill}}llll}
\hline
$j$ & $\mu$ & $\tr(\Box_s+\Box_t)/2N_c$ & $\tr(\Box_t-\Box_s)/N_c$ & 
$\tr L/N_c$ \\ \hline
\hline
0.00 & 0.00 & 0.4738(1) & -0.0002(1) & -0.0014(05) \\
     & 0.30 & 0.4739(3) &  0.0001(2) & -0.0008(14) \\
\hline
0.02 & 0.30 & 0.4742(3) & 0.0002(2) & -0.0012(15) \\
     & 0.35 & 0.4755(4) &  -0.0005(3) & 0.0005(24) \\
     & 0.50 & 0.4804(5) &  0.0024(2) & 0.0020(13) \\
     & 0.70 & 0.4773(3) &  0.0089(2) & 0.0127(11) \\
\hline
0.04 & 0.25 & 0.4735(3) & 0.0002(3) & -0.0026(14) \\
     & 0.30 & 0.4753(4) & 0.0007(4) & -0.0013(20) \\
     & 0.35 & 0.4764(4) & 0.0004(2) &  0.0003(13) \\
     & 0.40 & 0.4784(4) & 0.0012(3) &  0.0020(14) \\
     & 0.45 & 0.4792(3) & 0.0020(4) & -0.0016(18) \\
     & 0.50 & 0.4799(4) & 0.0025(3) &  0.0044(15) \\
     & 0.55 & 0.4793(4) & 0.0037(3) &  0.0004(17) \\
     & 0.60 & 0.4794(3) & 0.0051(3) &  0.0029(21) \\
     & 0.65 & 0.4778(3) & 0.0069(2) &  0.0065(11) \\
     & 0.70 & 0.4773(4) & 0.0092(3) &  0.0094(16) \\
     & 0.75 & 0.4752(2) & 0.0122(2) &  0.0188(19) \\
     & 0.80 & 0.4719(2) & 0.0172(3) &  0.0292(12) \\
     & 0.90 & 0.4648(3) & 0.0299(3) &  0.0719(17) \\
     & 1.00 & 0.4549(3) & 0.0442(3) &  0.1393(19) \\
     & 1.30 & 0.4277(4) & 0.0479(3) &  0.4286(20) \\
     & 1.40 & 0.4142(3) & 0.0259(4) &  0.3216(11) \\
     & 1.50 & 0.4060(2) & 0.0128(3) &  0.1019(13) \\
     & 1.60 & 0.4052(2) & 0.0114(3) &  0.0224(18) \\
     & 1.75 & 0.4052(2) & 0.0105(4) &  0.0018(18) \\
     & 2.50 & 0.4053(3) & 0.0115(4) &  0.0008(13) \\
\hline
0.06 & 0.30 & 0.4767(3) & 0.0003(2) & 0.0004(16) \\
     & 0.50 & 0.4791(4) &  0.0025(2) & -0.0035(14) \\
     & 0.70 & 0.4772(3) &  0.0095(4) & 0.0094(16) \\
\hline
\end{tabular*}
\caption{Raw data for gluonic observables}
\label{tab:gluon}
\end{table*}
\begin{table*}
\begin{tabular*}{0.9\textwidth}{l@{\extracolsep{\fill}}lllll}
\hline
$j$ & $\mu$ & $n_q$ & $\eq$ & $\pbp$ & $\qq$ \\ \hline
\hline
0.00 & 0.00 & -0.0004(04) & 0.3980(4) & 14.409(2) & -- \\
     & 0.30 & -0.0008(22) & 0.3994(22) & 14.399(7) & -- \\
\hline
0.02 & 0.30 & 0.0012(12) & 0.4028(16) & 14.389(5) & 0.0219(2) \\
     & 0.35 & 0.0030(24) & 0.4072(24) & 14.364(3) & 0.0293(3) \\
     & 0.50 & 0.0382(24) & 0.4496(40) & 14.254(10) & 0.0571(4) \\
     & 0.70 & 0.1506(32) & 0.5348(32) & 14.178(6) & 0.1099(5) \\
\hline
0.04 & 0.25 & 0.0036(14) & 0.4012(16) & 14.396(8) & 0.0343(1) \\
     & 0.30 & 0.0102(14) & 0.4126(16) & 14.360(4) & 0.0400(2) \\
     & 0.35 & 0.0144(12) & 0.4188(18) & 14.332(4) & 0.0469(3) \\
     & 0.40 & 0.0252(16) & 0.4358(16) & 14.292(8) & 0.0549(3) \\
     & 0.45 & 0.0332(24) & 0.4422(30) & 14.272(8) & 0.0634(5) \\
     & 0.50 & 0.0476(22) & 0.4558(22) & 14.244(8) & 0.0713(5) \\
     & 0.55 & 0.0648(30) & 0.4702(30) & 14.236(8) & 0.0817(5) \\
     & 0.60 & 0.0898(30) & 0.4902(30) & 14.196(8) & 0.0932(4) \\
     & 0.65 & 0.1238(26) & 0.5146(28) & 14.180(4) & 0.1084(4) \\
     & 0.70 & 0.1660(24) & 0.5476(26) & 14.156(4) & 0.1248(4) \\
     & 0.75 & 0.2302(34) & 0.5972(32) & 14.136(8) & 0.1461(5) \\
     & 0.80 & 0.3354(38) & 0.6850(32) & 14.084(8) & 0.1701(5) \\
     & 0.90 & 0.6664(56) & 0.9716(60) & 13.916(8) & 0.2222(9) \\
     & 1.00 & 1.2602(74) & 1.5168(72) & 13.560(8) & 0.2654(12) \\
     & 1.30 & 5.220(6) & 5.284(6) & 12.036(8) & 0.0956(3) \\
     & 1.40 & 7.040(4) & 7.060(4) & 11.312(4) & 0.0594(1) \\
     & 1.50 & 7.914(4) & 7.916(4) & 10.944(4) & 0.0346(3) \\
     & 1.60 & 7.996(3) & 7.996(3) & 10.916(4) & 0.0255(1) \\
     & 1.75 & 8.000(4) & 8.000(4) & 10.916(4) & 0.0207(1) \\
     & 2.50 & 8.000(2) & 8.000(2) & 10.924(4) & 0.0161(1) \\
\hline
0.06 & 0.30 & 0.0130(14) & 0.4184(14) & 14.331(6) & 0.0554(2) \\
     & 0.50 & 0.0530(22) & 0.4604(22) & 14.230(8) & 0.0855(8) \\
     & 0.70 & 0.1780(30) & 0.5574(30) & 14.131(8) & 0.1397(10) \\
\hline
\end{tabular*}
\caption{Raw data for quark observables (no zero subtraction applied
to $\eq$)}
\label{tab:quark}
\end{table*}

The most straightforward observable to analyse is the quark density
$n_q$, which as the timelike component of a conserved current requires
no renormalisation due to quantum corrections. There may, however,
still be lattice artefacts in both UV and IR r\'egimes, as illustrated
by the equivalent quantity for free fields on a $N_s^3\times N_t$
lattice:
\begin{equation}
n_{SB}^\latt(\mu)=\frac{4N_fN_c}{N_s^3N_t}\sum_k{
\frac{i\sin\kt_0\bigl[\sum_i\cos k_i-\frac{1}{2\kappa}\bigr]}
{\bigl[\frac{1}{2\kappa}-\sum_\nu\cos\kt_\nu\bigr]^2
 +\sum_\nu\sin^2\kt_\nu}}
\label{eq:nqfree}
\end{equation}
where
\begin{equation}
\kt_\nu = 
\begin{cases}
  k_0-i\mu = \frac{2\pi}{N_t}\bigl(n_0+\frac{1}{2}\bigr)-i\mu\,, & \nu=0,\\
  k_\nu = \frac{2\pi n_\nu}{N_s}\,, & \nu=1,2,3.
\end{cases}
\end{equation}
For free massless quarks
$\kappa=\frac{1}{8}$. 
In the large-$\mu$ limit $n_{SB}^\latt(\mu)$ saturates at a value $2N_fN_c$ 
per lattice site.
This is an artefact of non-zero lattice spacing, which we will discuss in more
detail in section~\ref{sec:saturate} below.
For $T=0$ the corresponding
continuum relation is (\ref{eq:nSB}).
\begin{figure}
\begin{center}
\includegraphics*[width=\colw]{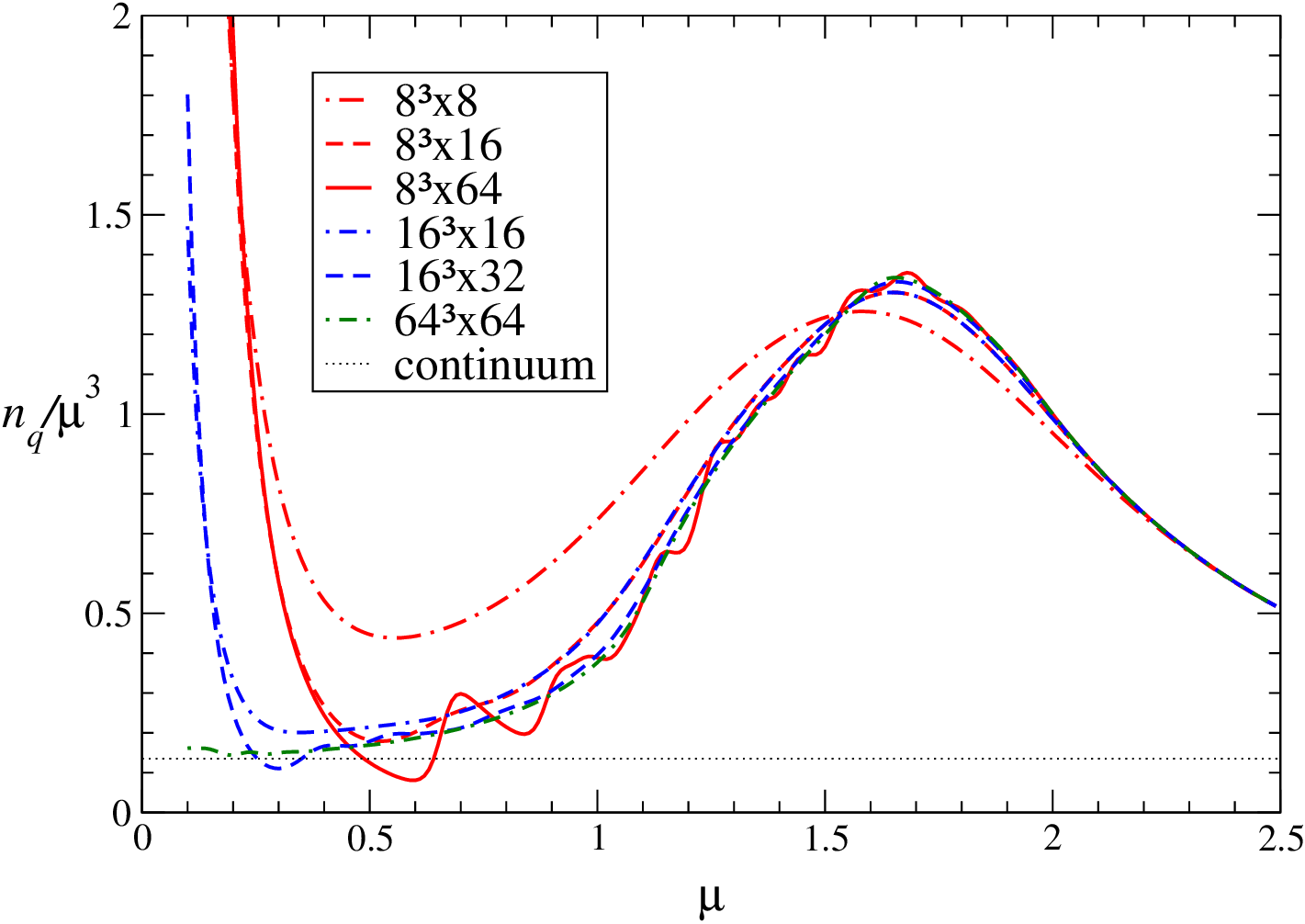}
\end{center}
\vspace{-3mm}
\caption{$n_{SB}^\latt/\mu^3$ for free massless Wilson fermions.}
\label{fig:n_qlatt}
\end{figure}
Fig.~\ref{fig:n_qlatt} 
plots $n_{SB}^\latt/\mu^3$ for several system sizes, and
shows there are significant departures from the continuum result both at 
small $\mu$ as a result of finite $N_s$, and at $\mu a\gtrsim\order(1)$ as 
a result of non-zero lattice spacing, species doubling etc. Of particular
interest are the pronounced wiggles seen especially on the $8^3\times64$ 
curve. These arise due to departures from sphericity of the Fermi surface in a 
finite spatial volume, and are visible whenever the temperature is much 
smaller than the mode spacing, i.e.\ $N_t\gg N_s$ \cite{Hands:2002mr}. 

\begin{figure}
\begin{center}
\includegraphics*[width=\colw]{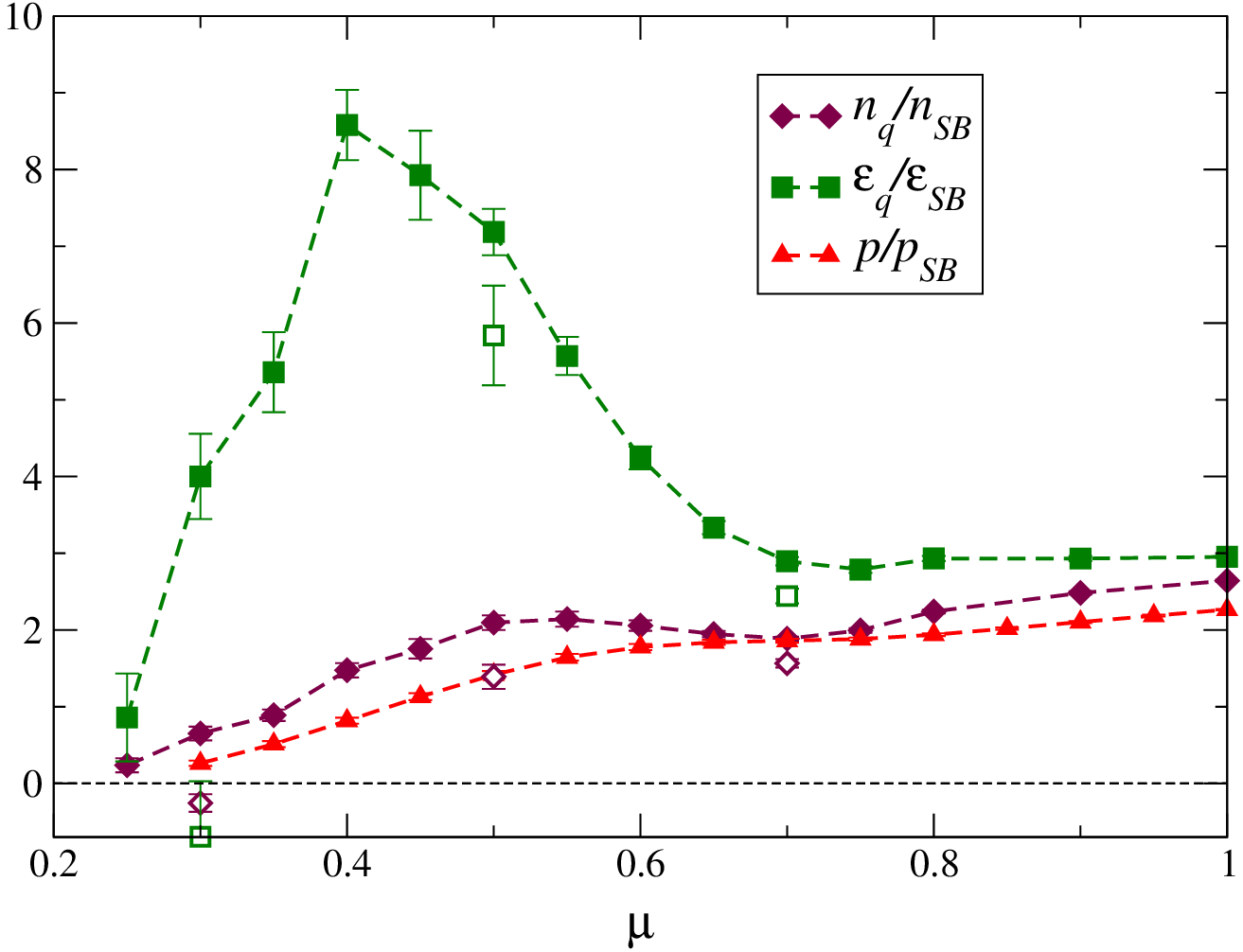}
\end{center}
\vspace{-3mm}
\caption{Ratio of thermodynamic observables to free field values versus
$\mu$ for $j=0.04$. Open symbols show extrapolations to $j=0$.}
\label{fig:eneps}
\end{figure}
The main implication in the window $\mu\in(0.25,1.0)$ where our analysis
is focussed 
is that $n_{SB}^\latt$ lies systematically above the continuum
value, with the ratio increasing for $\mu>0.5$.
To correct for this
lattice artefact, we have chosen to plot the ratio $n_q/n_{SB}^\latt$,
shown as a function of $\mu$ in fig.~\ref{fig:eneps}. The free field
density has been calculated for massless quarks, i.e.\ with
$\kappa=\frac{1}{8}$. In order to assess the systematic error due to
our lack of accurate knowledge of the non-zero quark mass, we repeated
the free field calculation with $\kappa=0.120$, corresponding to a
conservatively large $m_qa=0.167$, and found the resulting
$n_{SB}^\latt(\mu)$ to have a similar shape, with a maximum departure
over the $\mu$ range of interest of $O(20\%)$.  Where data are available we have
plotted as open symbols the result of linearly extrapolating $j\to0$; at
$\mu=0.3$ this yields $n_q\approx0$, consistent with being below the onset
expected at $\mu_o=m_\pi/2$. For $\mu=0.5,0.7$ the extrapolation results
in a downwards correction of $O$(20\%). We also 
note that for $\mu\gtrsim0.5$ the ratio is
roughly constant and greater than one, which is plausible if
Stefan--Boltzmann scaling sets in at large $\mu$, but with a Fermi
momentum $k_F>\mu$. Physically, this would result from degenerate
quark matter with a positive binding energy arising from
interactions. Note that a non-zero quark mass has the opposite effect,
tending to raise $\mu$ over $k_F$.

In fig.~\ref{fig:eneps}, we also plot the unrenormalised quark energy
density $\eq/\varepsilon_{SB}^\latt$ versus $\mu$, where
$\varepsilon_{SB}^\latt$ is evaluated using a formula similar to
(\ref{eq:nqfree}) and $\varepsilon_{SB}^\cont$ is given in
(\ref{eq:thermoSB}). Note that systematic errors due to non-zero quark
mass and incorrect vacuum subtraction are potentially larger in this
case, but have maximum impact at the lower end of the $\mu$ range of
interest. Once again, a $j\to0$ extrapolation has been done where possible, 
and the signal for $\mu=0.3$ is consistent with the vacuum.
As for quark number density, the ratio $\eq/\varepsilon_{SB}^\latt$
tends to a constant at large
$\mu$, which we interpret as evidence for the formation of a Fermi
surface. This limit is approached from above, however, and in the
range $0.35\lesssim\mu\lesssim0.5$ $\eq/\varepsilon_{SB}$ actually
peaks. The implications of this are discussed in section~\ref{sec:discussion}.

\begin{figure}
\begin{center}
\includegraphics*[width=\colw]{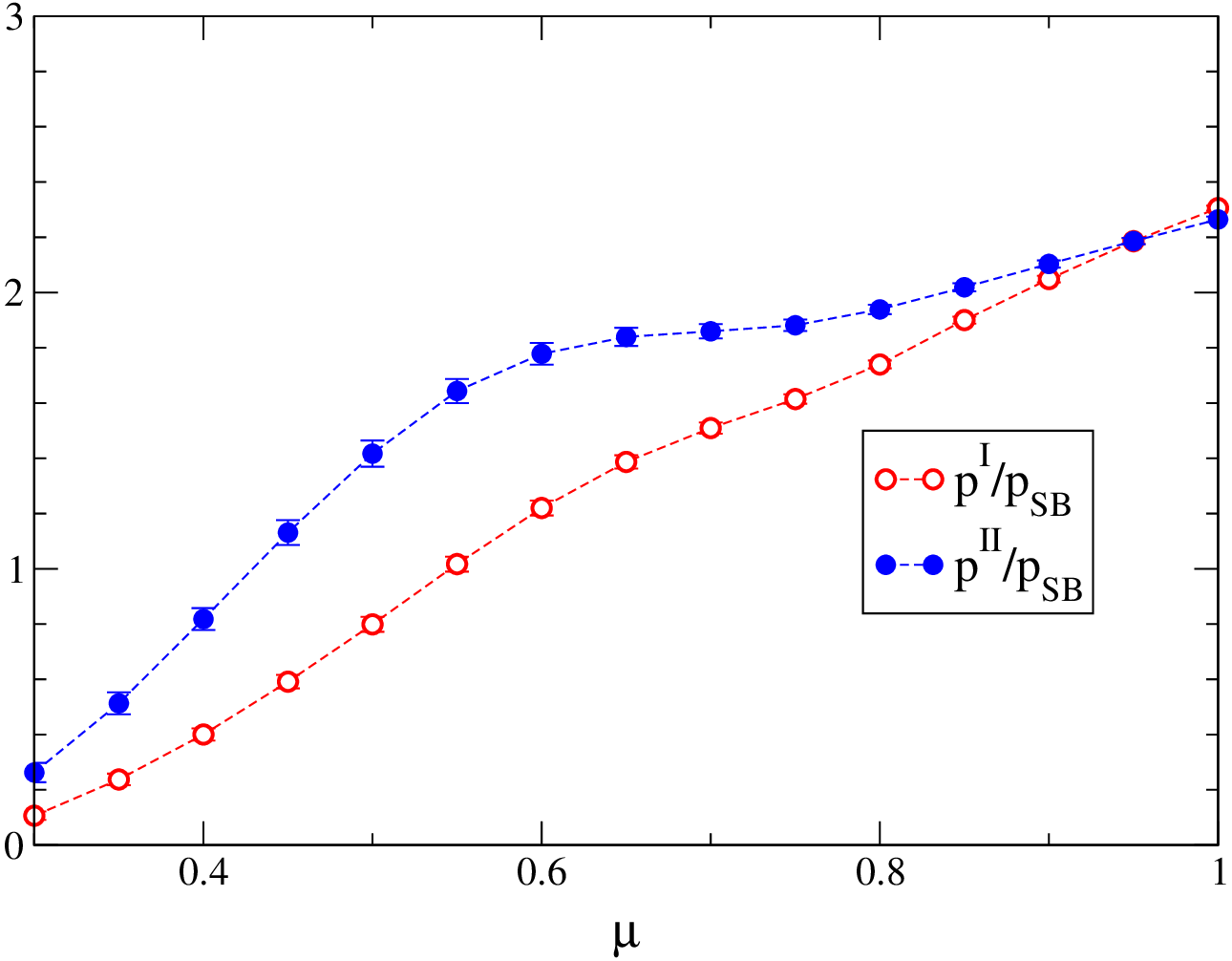}
\end{center}
\vspace{-3mm}
\caption{Pressure verus $\mu$ using two different integral methods}
\label{fig:pressure}
\end{figure}

The pressure $p$ is best calculated using the integral
method~\cite{Engels:1990vr}: i.e.\ $p(\mu)=\int_{\mu_o}^\mu
n_q(\mu^\prime)d\mu^\prime$. Note that since the only $\mu$-dependence
comes from the quark action, this expression gives in principle the
pressure of the entire system including both quarks and gluons.  In
practice for data taken away from both continuum and thermodynamic
limits we should make some attempt to correct for artefacts: we have
experimented with two different {\it ad hoc\/} procedures:
\begin{align}
\text{I}:\quad \frac{p}{p_{SB}}(\mu) &=
\frac{\int_{\mu_o}^\mu n_q(\mu^\prime)d\mu^\prime}
{\int_{\mu_o}^\mu n_{SB}^\latt(\mu^\prime)d\mu^\prime}\,;\\
\text{II}:\quad\frac{p}{p_{SB}}(\mu) &=
 \frac{\int_{\mu_o}^\mu \frac{n_{SB}^\cont}{n_{SB}^\latt}(\mu^\prime)
 n_q(\mu^\prime)d\mu^\prime}
 {\int_{\mu_o}^\mu n_{SB}^\cont(\mu^\prime)d\mu^\prime}\,,
\end{align}
where $n_{SB}^\cont$, $n_{SB}^\latt$ are defined in eqs. 
(\ref{eq:nSB},\ref{eq:nqfree}) respectively. Ultimately, data from different
physical lattice spacings will be required to determine which method is
preferred. Using an extended trapezoidal rule
to evaluate the integral on the $j=0.04$ data of fig.~\ref{fig:eneps}, 
we have estimated
both $p^I(\mu)$ and $p^{II}(\mu)$ and plot the results in 
fig.\ref{fig:pressure}.
In both cases the pressure rises monotonically from near zero at onset, but 
for method II there is some suggestion of a plateau in $p/p_{SB}$ for
$\mu\gtrsim0.6$. By $\mu\simeq1.0$ the two methods agree, with
$p\approx2p_{SB}$. This is consistent with the ratio $n_q/n_{SB}$ in the same
r\'egime, again suggesting a Fermi surface with
$k_F>\mu$. For comparison with the other fermionic
observables we have plotted $p^{II}/p_{SB}$ in fig.~\ref{fig:eneps}.

\begin{figure}
\begin{center}
\includegraphics*[width=\colw]{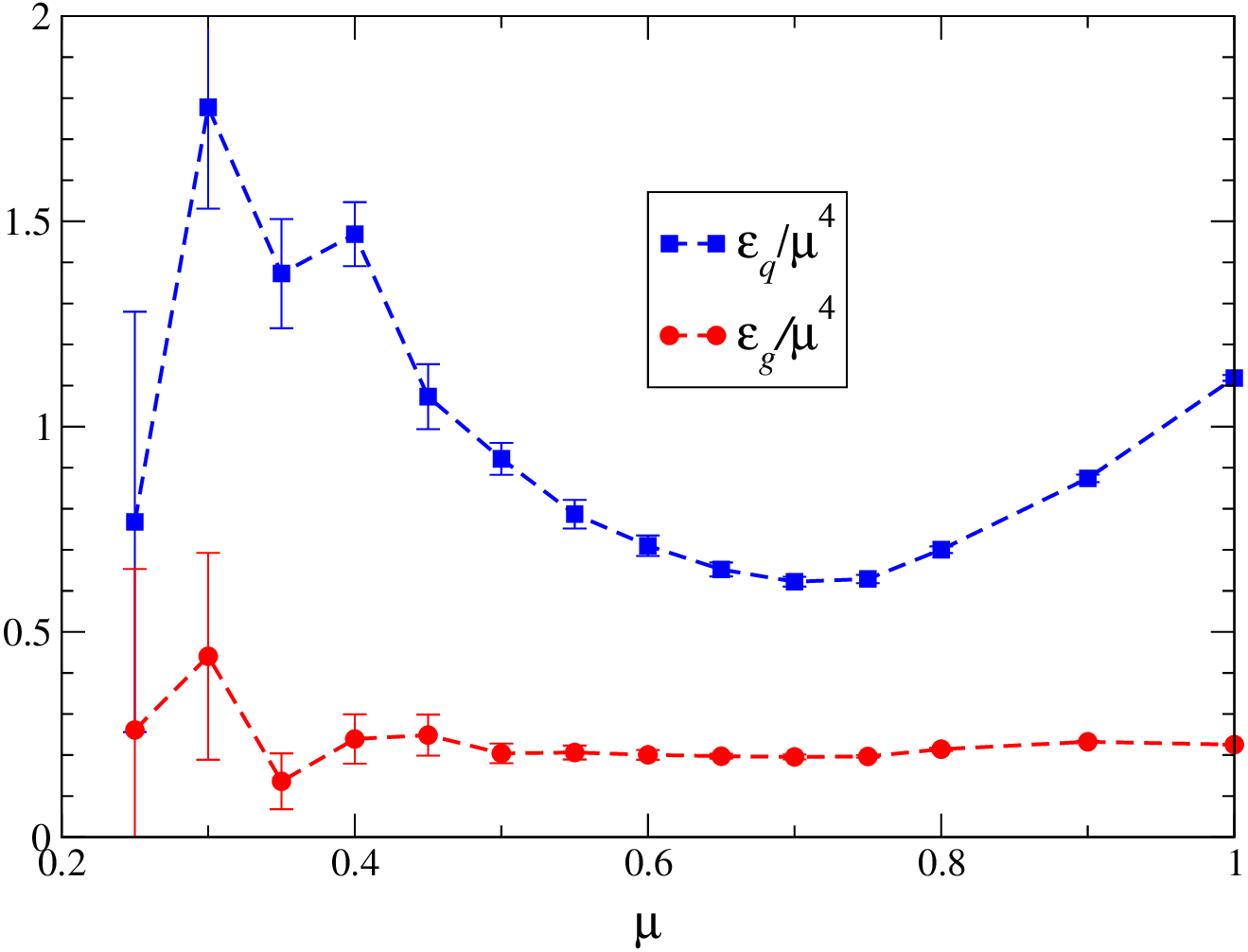}
\end{center}
\vspace{-3mm}
\caption{Comparison of bare quark and gluon energy densities versus $\mu$.}
\label{fig:energies}
\end{figure}

Next we turn to the gluon sector. Since at the current lattice spacing
the perturbative correction to the bare gluon energy density is likely
to be inadequate, in fig.~\ref{fig:energies} we content ourselves with
plotting unrenormalised data for both $\eg/\mu^4$ and for comparison
$\eq/\mu^4$,  at $j=0.04$. There is no evidence in our data for any significant
variation with $j$ for gluonic observables.
The plot illustrates: {\it(i)\/} the significant impact
of rescaling the fermionic data by $\varepsilon_{SB}^\latt$ (cf.\
fig.~\ref{fig:eneps}); {\it(ii)\/} for all $\mu>\mu_o$, the gluonic
contribution is a significant fraction of the total
energy density; {\it(iii)\/} most strikingly, $\eg$ scales as $\mu^4$
over the whole range of $\mu$.  On dimensional grounds, this is the
only physically sensible possibility if thermal effects are
negligible. We reiterate that gluonic contributions to thermodynamic
observables are not present in the SB formulae (\ref{eq:thermoSB}), so
that observations {\it(ii)\/} and {\it(iii)\/} are non-trivial
predictions of the simulation.  Note that our lack of knowledge about
the energy density renormalisation factors prevents us from making any
quantitative estimate of the gluon energy density as proportion of the
total energy density.

\begin{figure}
\begin{center}
\includegraphics*[width=\colw]{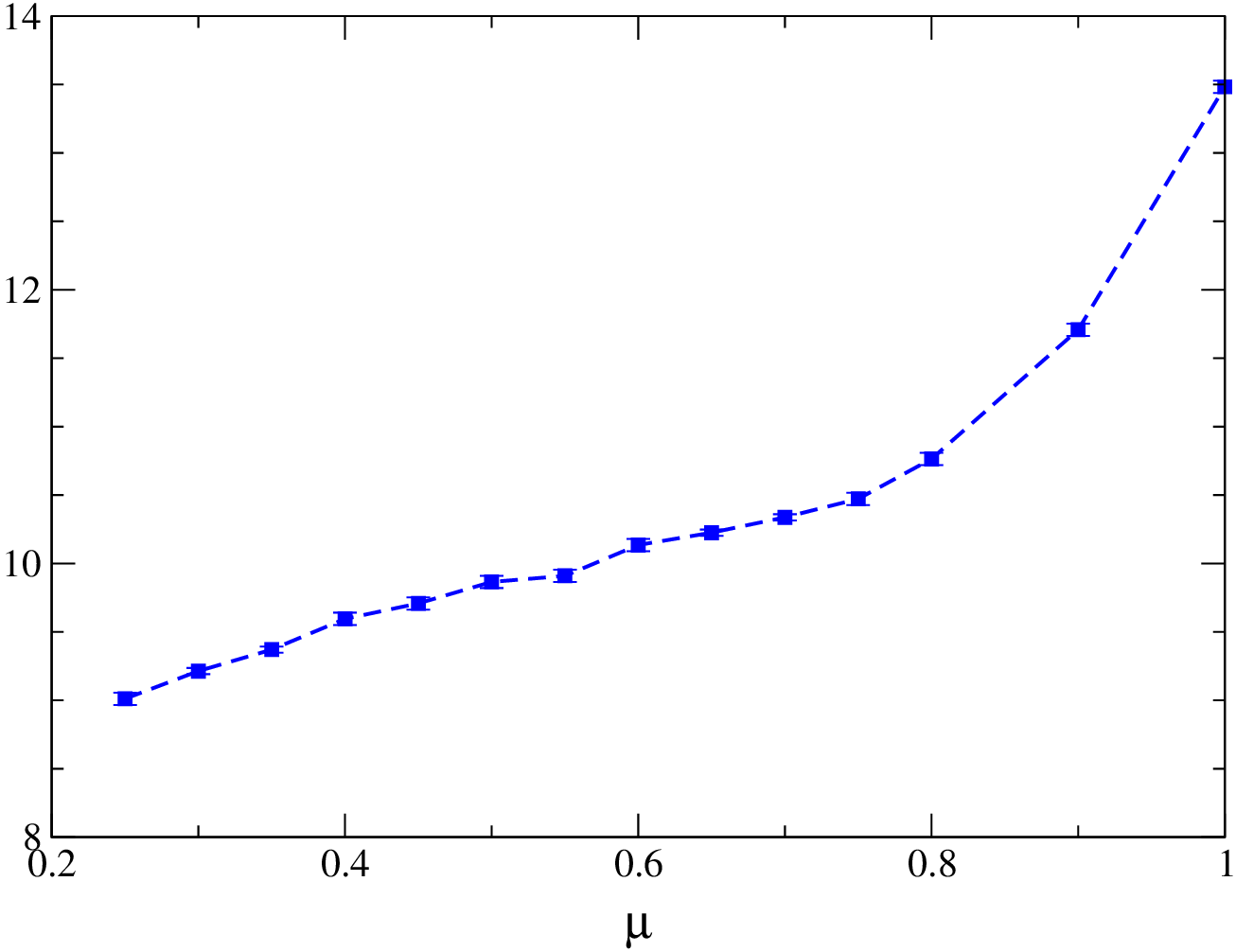}\\
\includegraphics*[width=\colw]{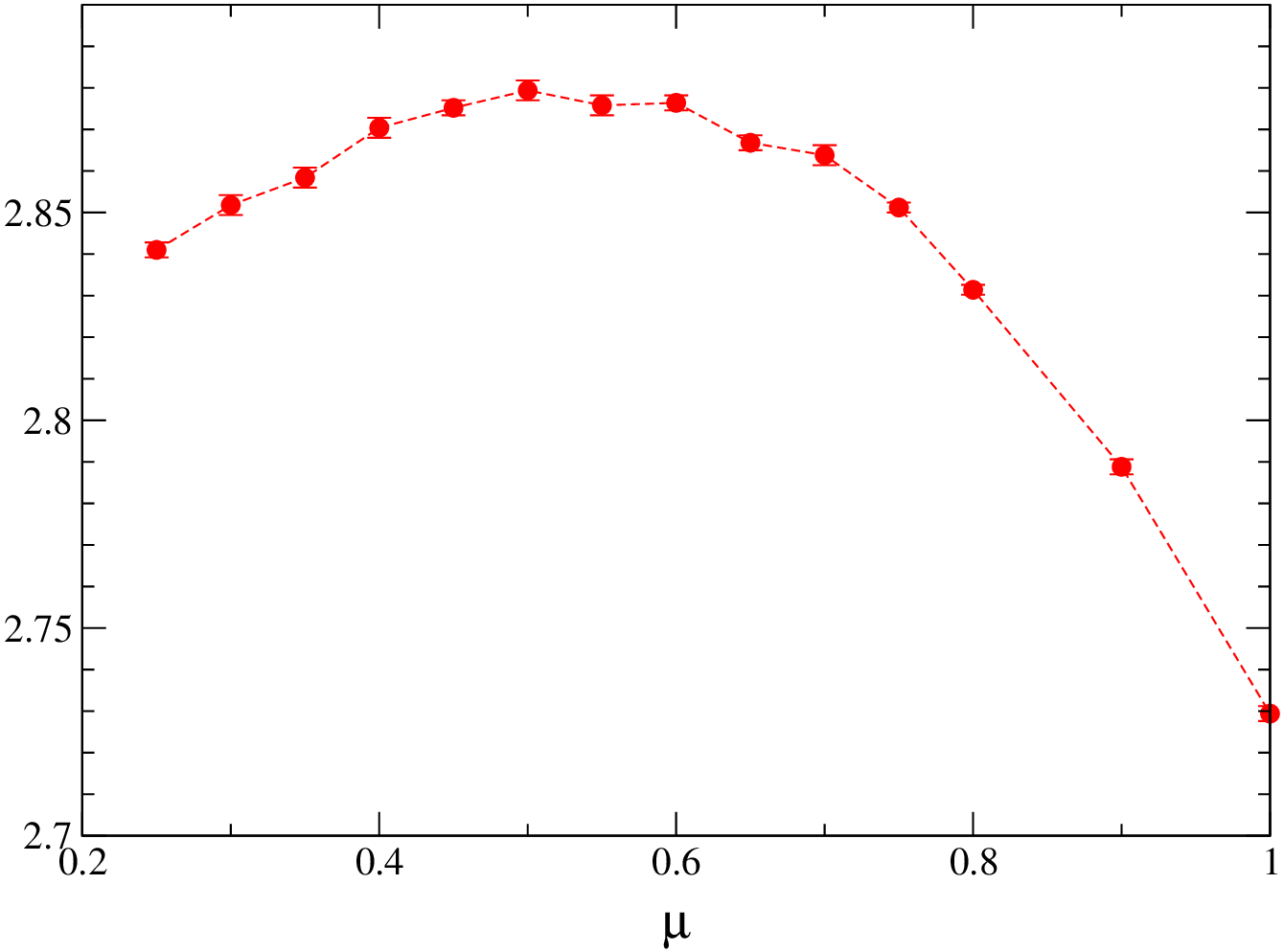}
\caption{Top: $\kappa^{-1}\tr(\One-\bra\bar\psi\psi\ket)$, and bottom:
  $3\beta/N_c\tr\bra \Box_t+\Box_s\ket$, as functions of $\mu$ at $j=0.04$.}
\label{fig:delta}
\end{center}
\end{figure}

The quark and gluon contributions to the conformal anomaly
(\ref{eq:conformq},\ref{eq:conformg}) are plotted in fig.~\ref{fig:delta}.
As for energy density, the raw data requires a
zero-$\mu$ subtraction.  To extract $\delta$, we also need information
on the lattice beta-functions, which require an extensive simulation
campaign and are not yet available. For this reason we restrict
ourselves to qualitative remarks. The quark contribution shown in the
upper panel of
fig.~\ref{fig:delta} increases monotonically from $\mu=0$. The sign
of $\partial\kappa/\partial a$ is found to be negative in QCD simulations
\cite{AliKhan:2001ek,Gockeler:2004rp} (although it can change sign as
the chiral limit $\kappa\to\kappa_c$ is approached away from the
continuum limit \cite{AliKhan:2001ek}). This is in accord with what we
have found in our SU(2) simulations at various $\beta,\kappa$
\cwilsa. We conclude that the quark
contribution to $\delta$ is negative.  The gluon contribution, by
contrast, starts positive for $\mu_o\lesssim\mu\lesssim0.6$ before
changing sign to decrease montonically for $\mu>0.6$ (recall
$\partial\beta/\partial a<0$ due to asymptotic freedom).

In fact, we expect the gluon contribution to dominate, since
$\delta\kappa/\delta\beta\approx \order(0.01)$ along lines of constant physics
\cite{Gockeler:2004rp}. This is clearly required for consistency with
fig.~\ref{fig:eneps}, since the sign of $\delta$ must coincide with the sign of
$\varepsilon/\varepsilon_{SB}-p/p_{SB}$, which looks positive for
$0.4\lesssim\mu\lesssim0.6$. We also get a clue about the unknown
renormalisation factor for $\eq$, since if $\delta$ is to go negative
at large $\mu$, $\varepsilon_{\text{ren}}/\varepsilon_{SB}<p/p_{SB}$ in this r\'egime.

The non-monotonic behaviour of the plaquette has been predicted in a $\chi$PT
study in which asymptotic freedom of the gauge coupling is taken into account
\cite{Metlitski:2005db}, and has also been observed in
recent simulations with staggered fermions \cite{Alles:2006ea}. It
can be understood in terms of Pauli blocking.  In quark matter as
$\mu$ increases, the number of $q\bar q$ pairs available for vacuum
polarisation corrections to the gluon decreases, since only states
close to the Fermi surface can participate. In the limit of complete
saturation (i.e.\ one quark of each color, flavor and spin per lattice
site) the gluon dynamics hence resembles that of the quenched theory, so
$\tr\Box(\mu\to\infty)<\tr\Box(\mu=0)$. 
Assuming a smooth passage to the
limit we deduce $\delta<0$ at large $\mu$.

To summarise the thermodynamic information, we have been able to
extract $n_q(\mu)$ and $p(\mu)$ directly from the simulation --- any
remaining UV and IR artefacts in fig.~\ref{fig:eneps} can be
controlled by simulations closer to thermodynamic and continuum
limits. The status of $\eq,\eg$ is less secure, because these
quantities require renormalisation by an as yet undetermined factor;
similarly the trace of the energy-momentum tensor $\delta$ requires
knowledge of the lattice beta-functions. In each case, though, the
rescaling factor is $\mu$-independent, so the shape of the data in
figs.~\ref{fig:eneps} and~\ref{fig:delta} is correct.  However, the
behaviour of $\varepsilon_q$ in fig.~\ref{fig:eneps}, and $\delta_g$
in fig.~\ref{fig:delta} give a strong hint of two qualitatively
distinct high density regions: {\it(i)\/} for
$\mu_o\lesssim\mu\lesssim0.65$ $\eq\approx O(5\varepsilon_{SB})$, and
$\delta_g>0$; {\it(ii)\/} for $\mu\gtrsim0.65$
$\eq/\varepsilon_{SB}\approx n_q/n_{SB}$, $\delta_g<0$. All
thermodynamic quantities seem to have the same $\mu$-scaling as their
Stefan--Boltzmann counterparts in this higher density r\'egime, although
$n_q/n_{SB}\approx p/p_{SB}\approx2$.

\subsection{The approach to saturation}
\label{sec:saturate}

\begin{figure}
\begin{center}
\includegraphics*[width=\colw]{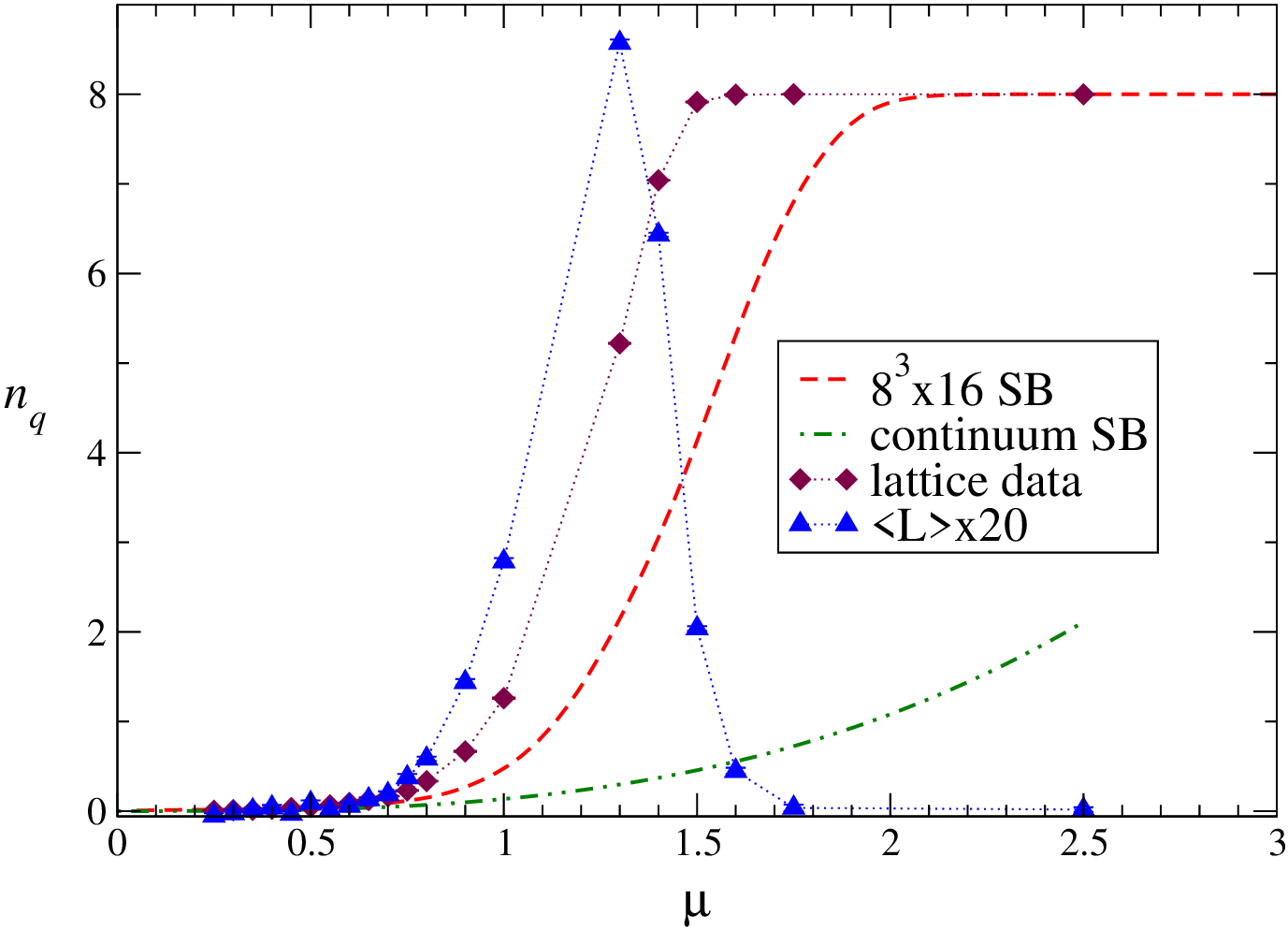}
\caption{Quark number density $n_q$ and Polyakov loop $L$ vs. $\mu$ for values
up to $\mu=2.5$.  All lattice data are at $j=0.04$.}
\label{fig:saturate}
\end{center}
\end{figure}
In figure~\ref{fig:saturate} we plot $n_q(\mu)$ including values of $\mu$ up to
2.5, as tabulated in tables~\ref{tab:params},~\ref{tab:gluon} and 
\ref{tab:quark}, together with both continuum (\ref{eq:nSB}) and
lattice (\ref{eq:nqfree}) expectations for non-interacting quarks.
On any system with a non-zero lattice spacing, there is a point at which every
lattice site is occupied by the maximum value $2N_cN_f$ allowed by the Pauli
exclusion principle. For free Wilson fermions this occurs for $\mu
a\approx2.0$; for the interacting quarks studied here the saturation threshold
drops to $\mu a \approx1.5$. We also see that the relation
$n_q\approx2n_{SB}^{\rm lat}$ continues to hold all the way up to the threshold;
there is no sign of asymptotic freedom.  In this respect, the
situation bears some similarity to that of QCD at temperatures between $T_c$ and
$3T_c$, where lattice simulations have also uncovered a deconfined,
but still strongly interacting system.  The nature of this system is
quite different, however, as in the high-temperature case a slow
approach towards Stefan--Boltzmann predictions is observed, while in
the present case the strong binding energy remains unchanged in the
entire domain studied.

As discussed above, in a saturated system, virtual $q\bar q$ 
pairs are suppressed, leading to the expectation that the gluodynamics should be
that of the quenched theory. In fact, inspection of table~\ref{tab:gluon} shows
this is a little simplistic, since $\varepsilon_g$ remains non-vanishing 
even once saturation is complete. Strong coupling considerations suggest that
the saturated system has a quark-induced effective action that can be
expanded in even powers of the Polyakov loops.
The resulting $S_{\text{eff}}$ distinguishes between $\Box_s$ and
$\Box_t$ hence yielding $\varepsilon_g\not=0$, but respects the global
$Z_2$ centre symmetry of the quenched action, consistent with
$\langle\tr L\rangle=0$.

Just below saturation, 
$\tr L$ rises rapidly. In this r\'egime the theory
resembles a p-type semiconductor, in that the low energy excitations are 
de-confined holes.
The holes appear weakly interacting: $\langle qq\rangle$, which 
also measures the density of hole-hole pairs, is small, and
there is little evidence from
$N_{cg}$ (table~\ref{tab:params}) for a light bound state.

\subsection{Order parameters}
\label{sec:orderps}

\begin{figure}
\begin{center}
\includegraphics*[width=\colw]{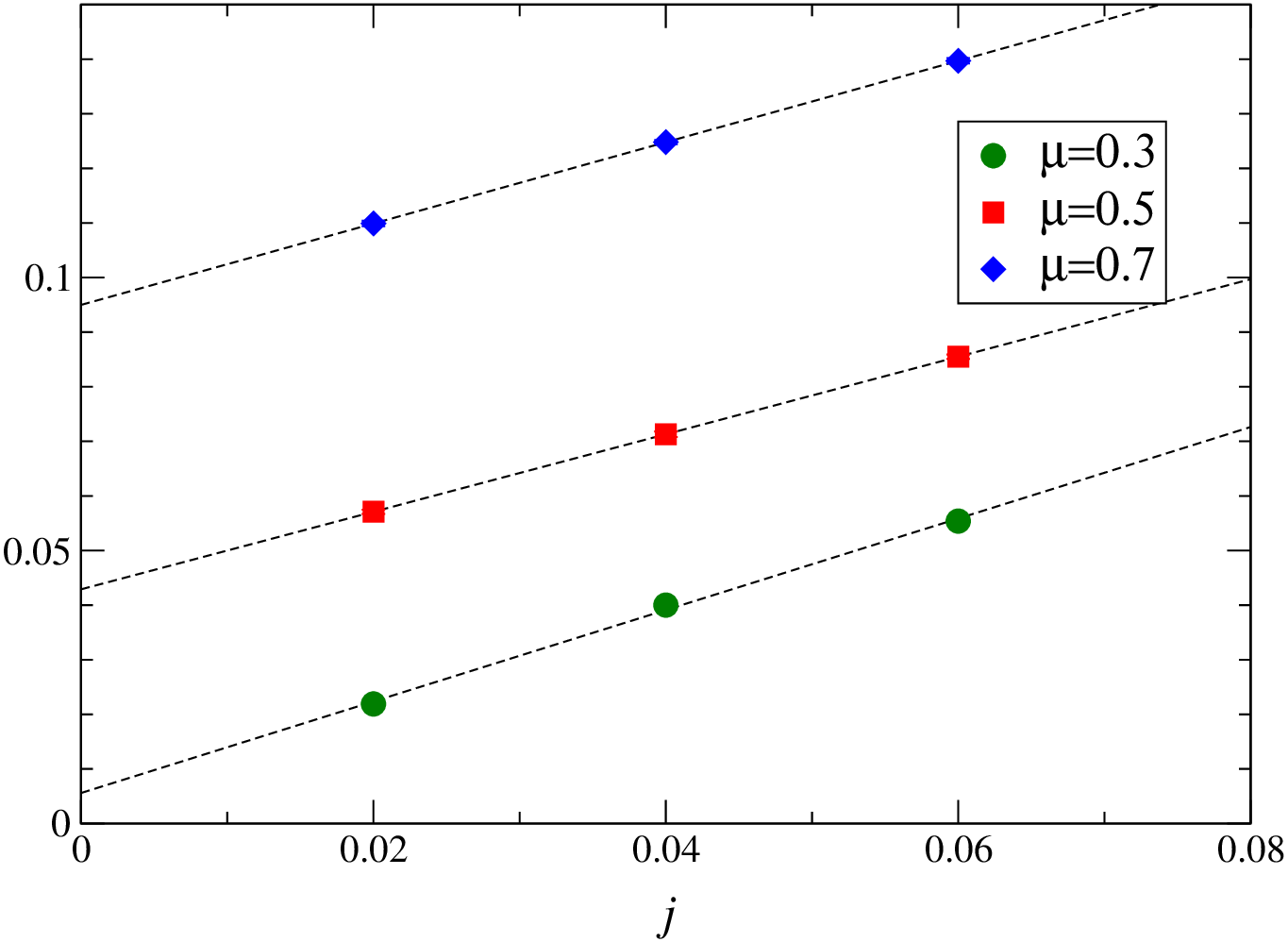}
\end{center}
\vspace{-3mm}
\caption{Superfluid condensate $\qq$ vs. $j$ for various $\mu$}
\label{fig:varyj}
\end{figure}

Might it be possible to reconcile the behaviour reported in
section~\ref{sec:thermo}
with the models of section~\ref{sec:models}? To elucidate this question
we next review order
parameters for superfluidity and deconfinement.
\begin{figure}
\includegraphics*[width=\colw]{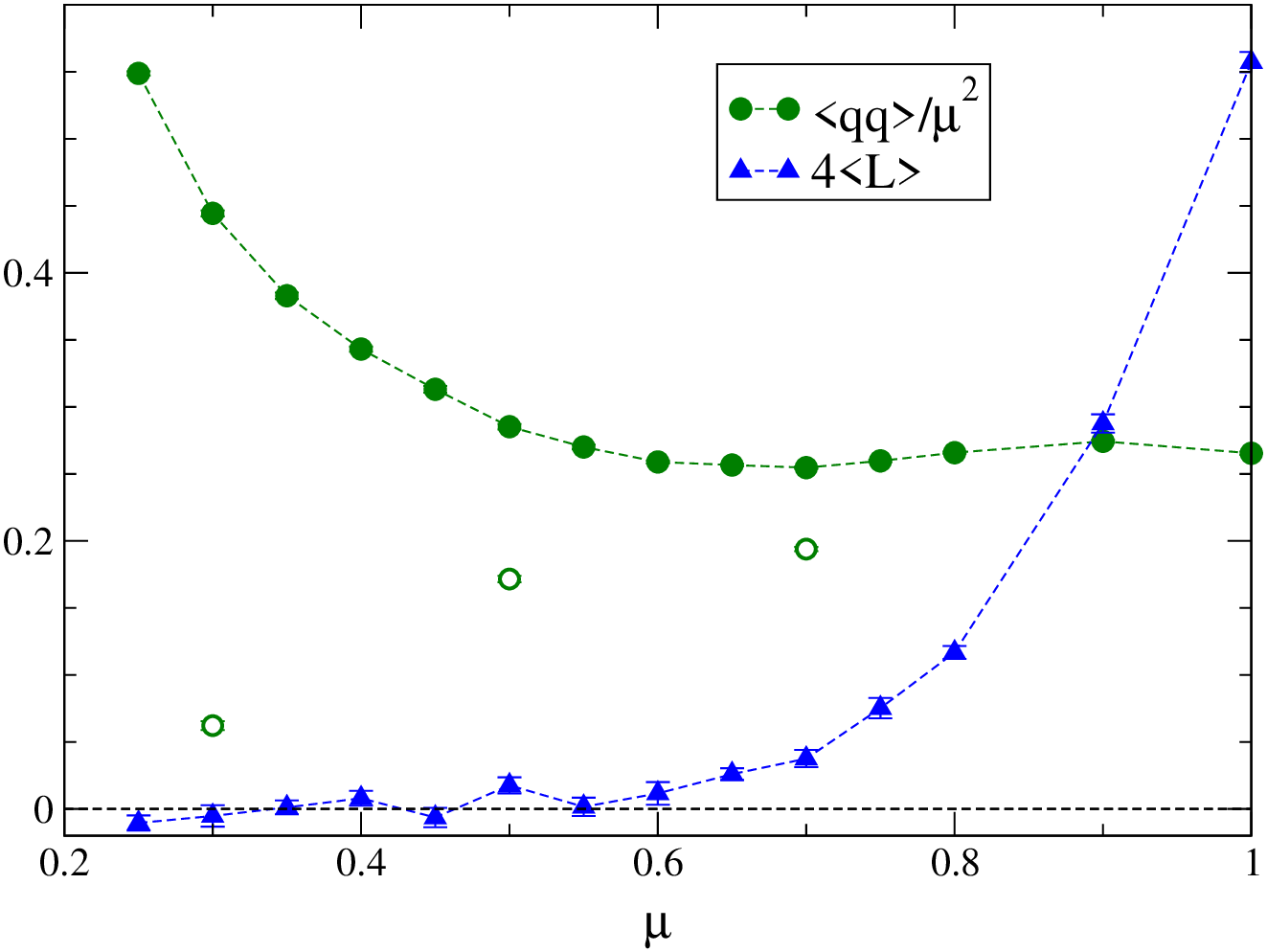}
\caption{The Polyakov loop $L$ and the superfluid diquark condensate
  $\qq/\mu^2$ for $j=0.04$ as function of $\mu$. Open symbols show 
$\qq/\mu^2$ extrapolated to $j=0$.}
\label{fig:diq-deconf}
\end{figure}
In fig.~\ref{fig:diq-deconf} we plot both the superfluid condensate 
$\qq$ given in (\ref{eq:qq}), rescaled by a factor $\mu^{-2}$, 
and the Polyakov loop $L$. Both show a marked change of behaviour at
$\mu\simeq0.6$; for $\mu$ greater than this value $\qq/\mu^2$
is approximately constant, while $L$ rises from zero. To interpret the
superfluid condensate, we first need to compare data taken at varying $j$, 
shown in fig.~\ref{fig:varyj} together with a linear extrapolation to $j=0$.
The slight non-linearity of the $\mu=0.3$ data 
suggests that below onset,
$\lim_{j\to0}\qq=0$, whereas for $\mu=0.5$, 0.7 the data extrapolate
to a non-vanishing intercept, implying a non-vanishing condensate
and hence spontaneous breaking of baryon number symmetry,
i.e.\ superfluidity. Extrapolated values are plotted as open symbols in
fig.~\ref{fig:diq-deconf}.

Now, in the Bose condensate phase expected for
$\mu_o<\mu<\mu_d$, $\chi$PT predicts \ckstvz
\begin{equation}
\frac{\qq_{\chi PT}}{\mu^2}
\propto\frac{1}{\mu^2}\sqrt{1-\frac{\mu_o^4}{\mu^4}},
\end{equation}
which is a monotonically decreasing function for $\mu\geq\root 4 \of {2}\mu_o$.
For a Fermi surface perturbed by a weak $qq$ attractive force, by contrast,
the order parameter counts the number of Cooper pairs condensed in
the ground state, which all originate in a shell of thickness $\Delta$ around
the Fermi surface, where $\Delta$ is the superfluid gap: hence
\begin{equation}
\frac{\qq_{BCS}}{\mu^2}\propto \Delta.
\end{equation}
The data of fig.~\ref{fig:diq-deconf} support this scenario for
$\mu\gtrsim0.6$, with $\Delta$ independent of $\mu$. Further support
for the importance of quark degrees of freedom at large $\mu$ comes
from the Polyakov loop, which rises smoothly from zero at
$\mu\simeq0.65$. We thus tentatively assign $\mu_d\simeq0.65$, marking
a transition from confined scalar ``nuclear'' matter to deconfined
quark matter. In condensed matter physics parlance this transition
would be characterised as one from BEC to BCS. Exposing the detailed nature of
the transition will require many more simulations using a variety of
source strengths, lattice spacings, and spatial volumes.

\subsection{Static quark potential}
\label{sec:potential}

\begin{figure}
\includegraphics*[width=\colw]{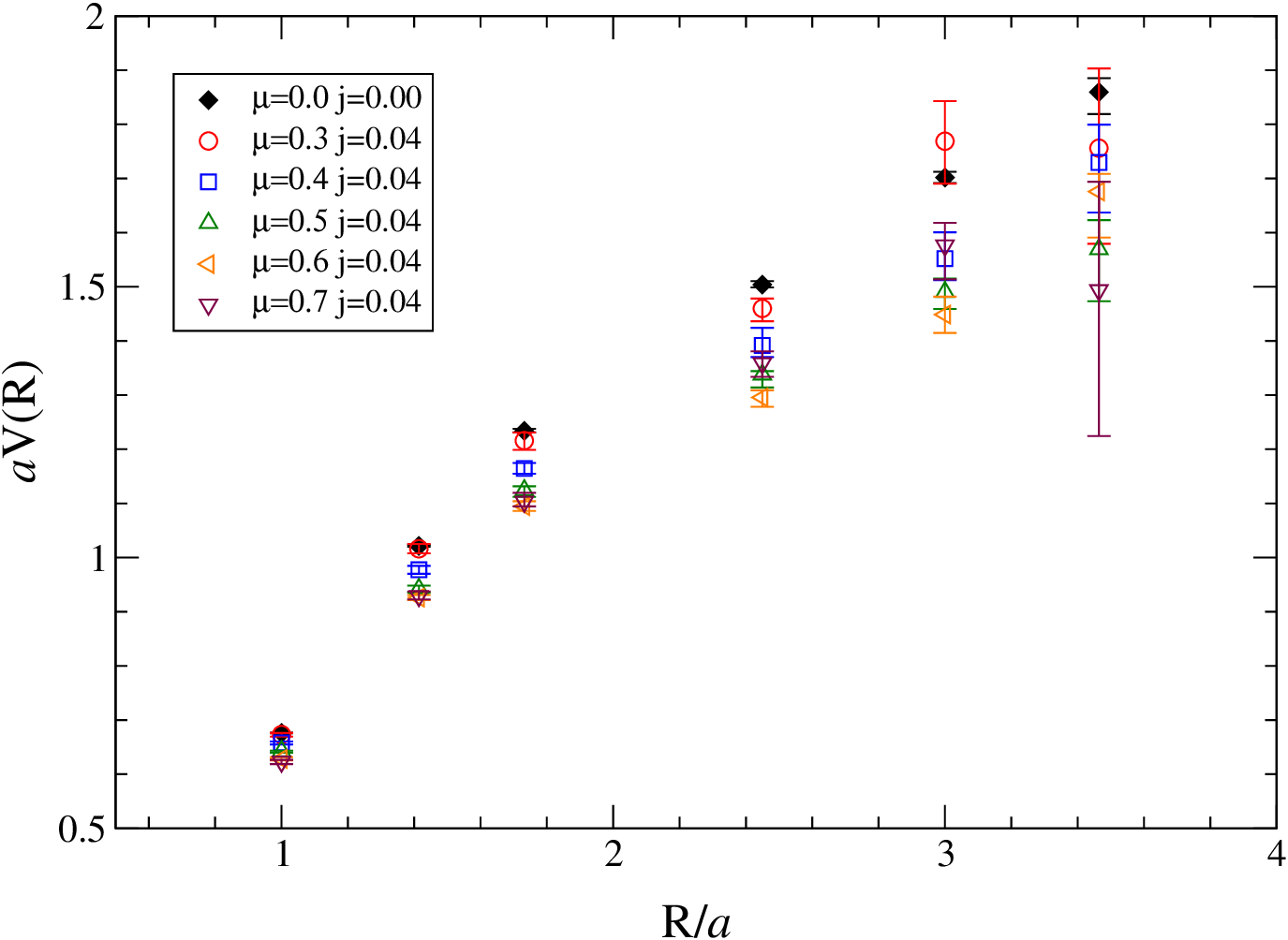}
\caption{The static quark potential for various values of the chemical
  potential $\mu$ and diquark source $j=0.04$, together with the
  $\mu=j=0$ potential shown in fig.~\protect\ref{fig:V0}.}
\label{fig:Vmu}
\end{figure}

\begin{figure}
\includegraphics*[width=\colw]{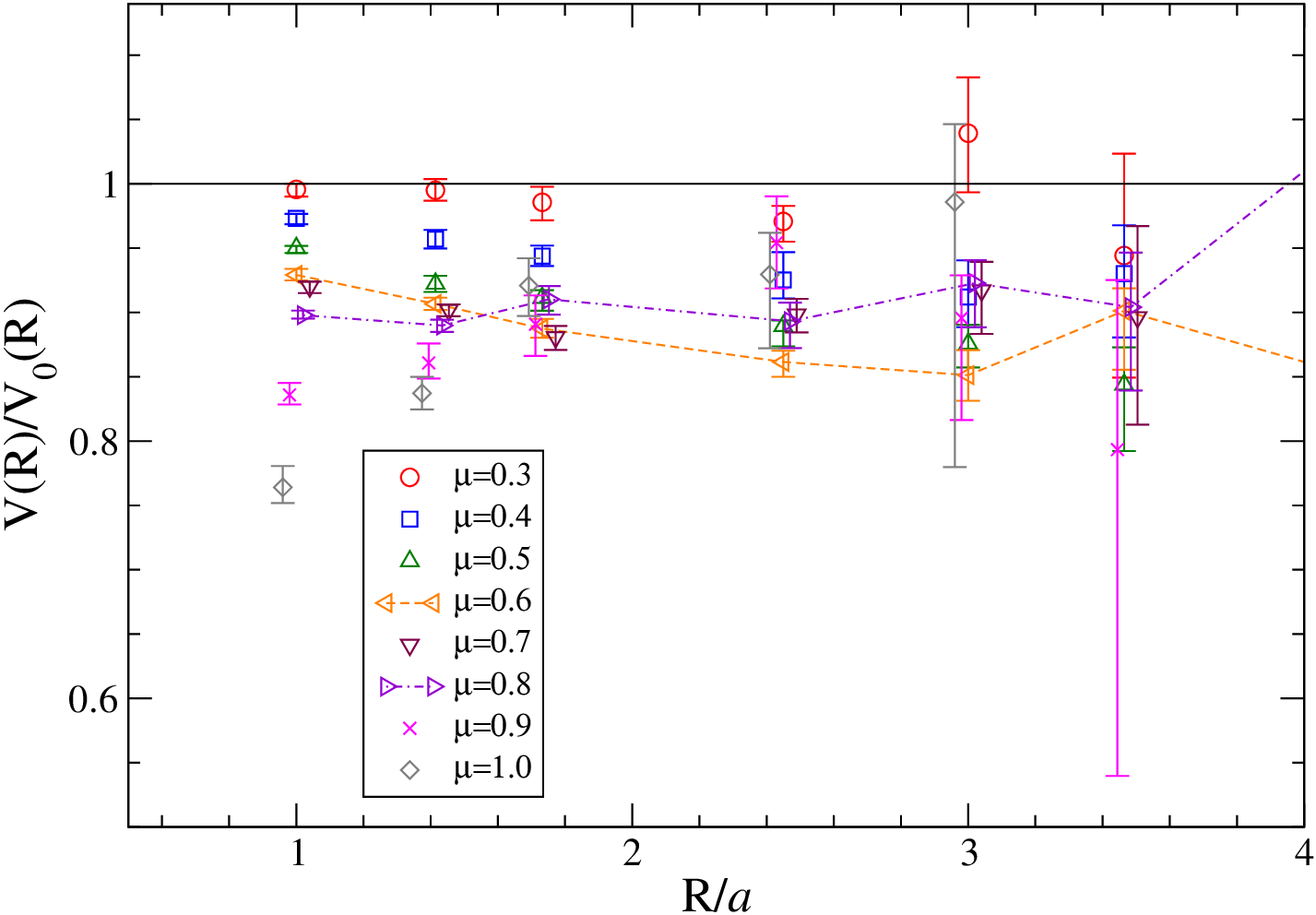}
\caption{The static quark potential for various values of the chemical
  potential $\mu$ and diquark source $j=0.04$, divided the
  $\mu=j=0$ potential shown in fig.~\protect\ref{fig:V0}.}
\label{fig:Vscale}
\end{figure}

The screening effect of the dense medium can be further investigated
by studying how the static quark potential changes as the chemical
potential increases.
In figure~\ref{fig:Vmu} we show the static quark potential for various
values of $\mu$ and $j=0.04$.  Up to $\mu=0.3$ there is no change
from the $\mu=0$ potential, while in the intermediate phase we see
clear evidence of screening due to a nonzero baryon density.  This is
even clearer in fig.~\ref{fig:Vscale}, where the $\mu=j=0$ potential
has been factored out.

\begin{figure}
\includegraphics*[width=\colw]{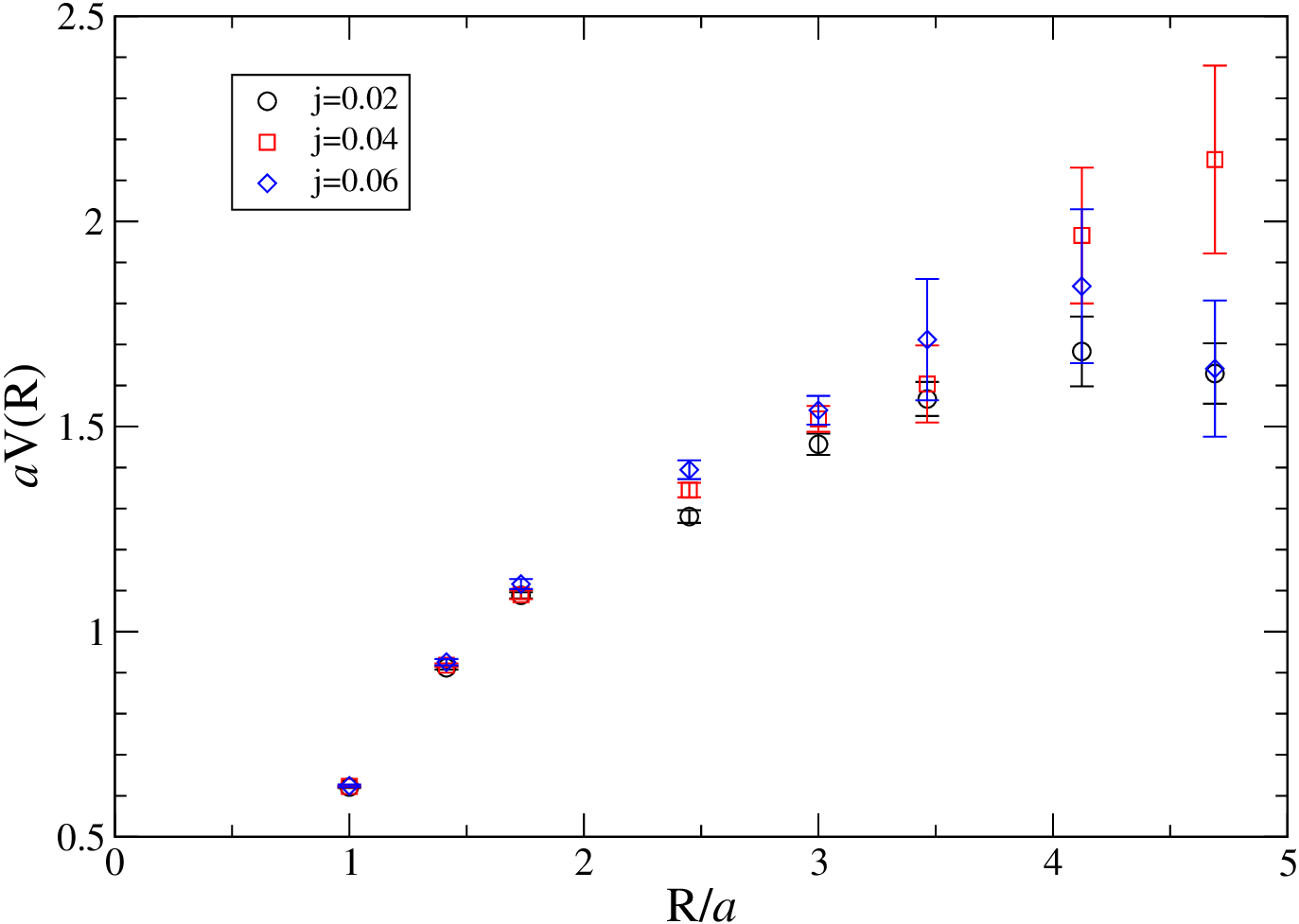}
\caption{The static quark potential at $\mu=0.7$ for different values
  of $j$.}
\label{fig:V_jvary}
\end{figure}

In the deconfined phase a new pattern emerges, where the short
distance potential is strongly modified, while at long distances we
appear to see an increase with increasing $\mu$ rather than a decrease
as expected.  There are however indications that the long-distance
screening at $\mu=0.7$ may become slightly stronger as $j\to0$, as
fig.~\ref{fig:V_jvary} shows.  No dependence on $j$ is seen at
$\mu=0.3$ or $\mu=0.5$.  We do not have an understanding of these
effects at present, although it is possible that lattice artefacts may
contribute to the short-distance modifications.  This can be resolved
by going to finer lattices.  Likewise, larger lattices will be
required to determine the long distance potential to greater accuracy.

\subsection{Gluon propagator}
\label{sec:gluon}

We have computed the gluon propagator by fixing the configurations to
Landau gauge using an overrelaxation algorithm to a precision
$|\partial_\nu A_\nu|^2<10^{-10}$.  The approach and notation is
analogous to that of \cite{Leinweber:1998uu}. At nonzero chemical
potential, which defines a preferred rest frame, the gluon propagator
can be decomposed as
\begin{align}
\begin{split}
D_{\mu\nu}(q) &= P^M_{\mu\nu}(q)D_M(q_0^2,\vec{q}^2)\\
&\phantom{=} + P^E_{\mu\nu}(q)D_E(q_0^2,\vec{q}^2) +
 P^L_{\mu\nu}(q)D_L(q_0^2,\vec{q}^2)\,,
\end{split}
\label{eq:gluonstruct}\\
\intertext{where}
P^M_{ij}(q) &= (\delta_{ij}-\frac{q_iq_j}{\vec{q}^2})\,; \quad
P^M_{00}(q) = P^M_{i0} = 0\,;\\
P^E_{\mu\nu}(q) &=
\delta_{\mu\nu}-\frac{q_{\mu}q_{\nu}}{q_0^2+\vec{q}^2}
 - P^M_{\mu\nu}(q)\,;\\
P^L_{\mu\nu}(q) &= \frac{q_{\mu}q_{\nu}}{q_0^2+\vec{q}^2}\,;\\
q_\nu &= \frac{2}{a}\sin\Bigl(\frac{\pi n_\nu}{L_\nu}\Bigr), \quad
n_\nu=-\frac{L_\nu}{2}+1,\ldots, \frac{L_\nu}{2} \,.
\end{align}
$D_M$ is the magnetic (spatially transverse) gluon propagator
and $D_E$ is the electric (spatially longitudinal) propagator, while
the longitudinal propagator $D_L$ is zero in Landau gauge.

\begin{figure}
\includegraphics*[width=\colw]{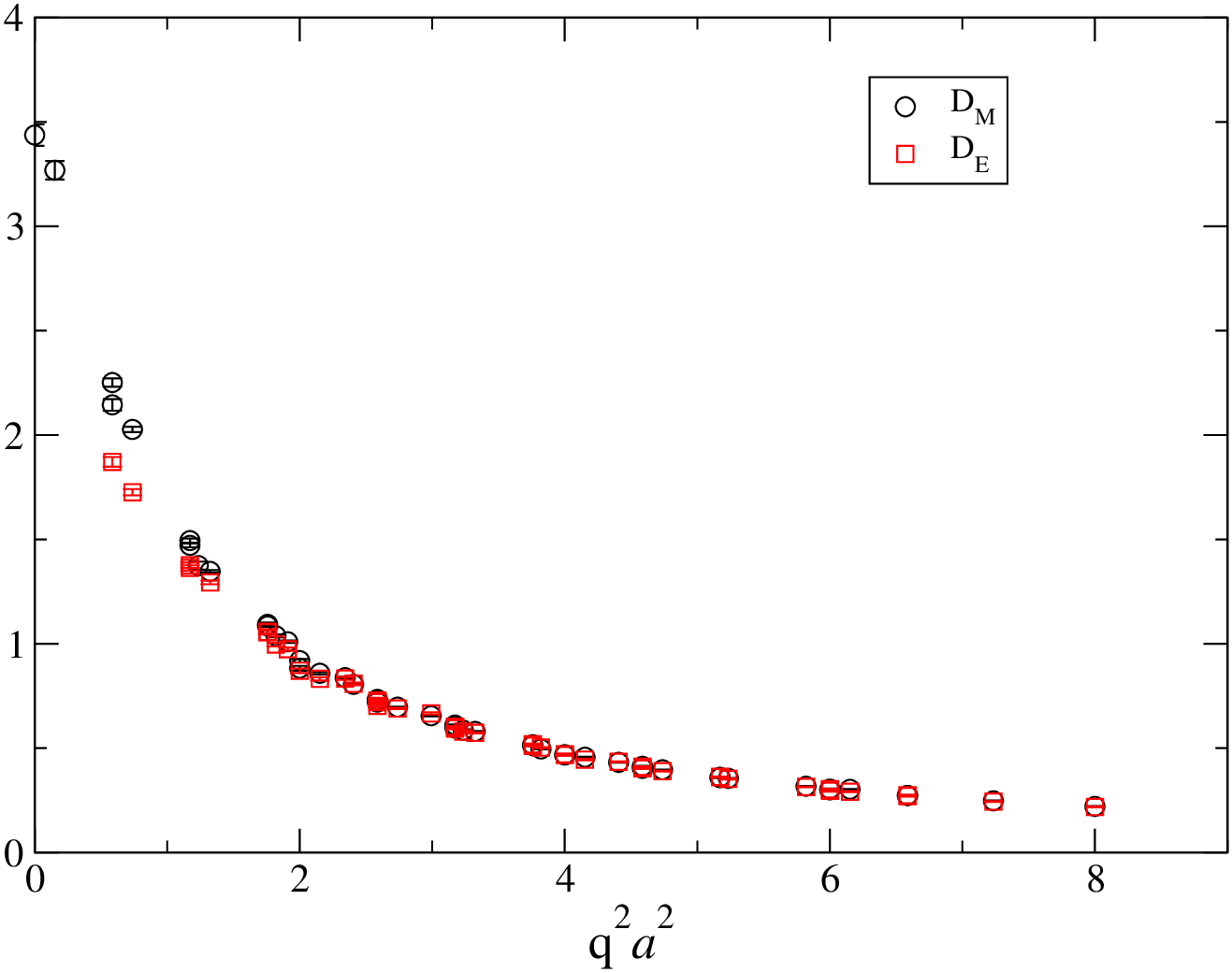}
\caption{Gluon propagator at $\mu=j=0$.}
\label{fig:gluon0}
\end{figure}

Figure \ref{fig:gluon0} shows the gluon propagator at $\mu=j=0$ as a
function of 4-momentum $q^2=q_0^2+\vec{q}^2$.  Since the Lorentz
symmetry remains unbroken here there is only one form factor, and the
small splitting between electric and magnetic gluon is most likely a
finite volume effect, similar to the asymmetric finite volume
effects on the tensor structure observed in the SU(3) gluon propagator
\cite{Leinweber:1998uu}.

\begin{figure}
\includegraphics*[width=\colw]{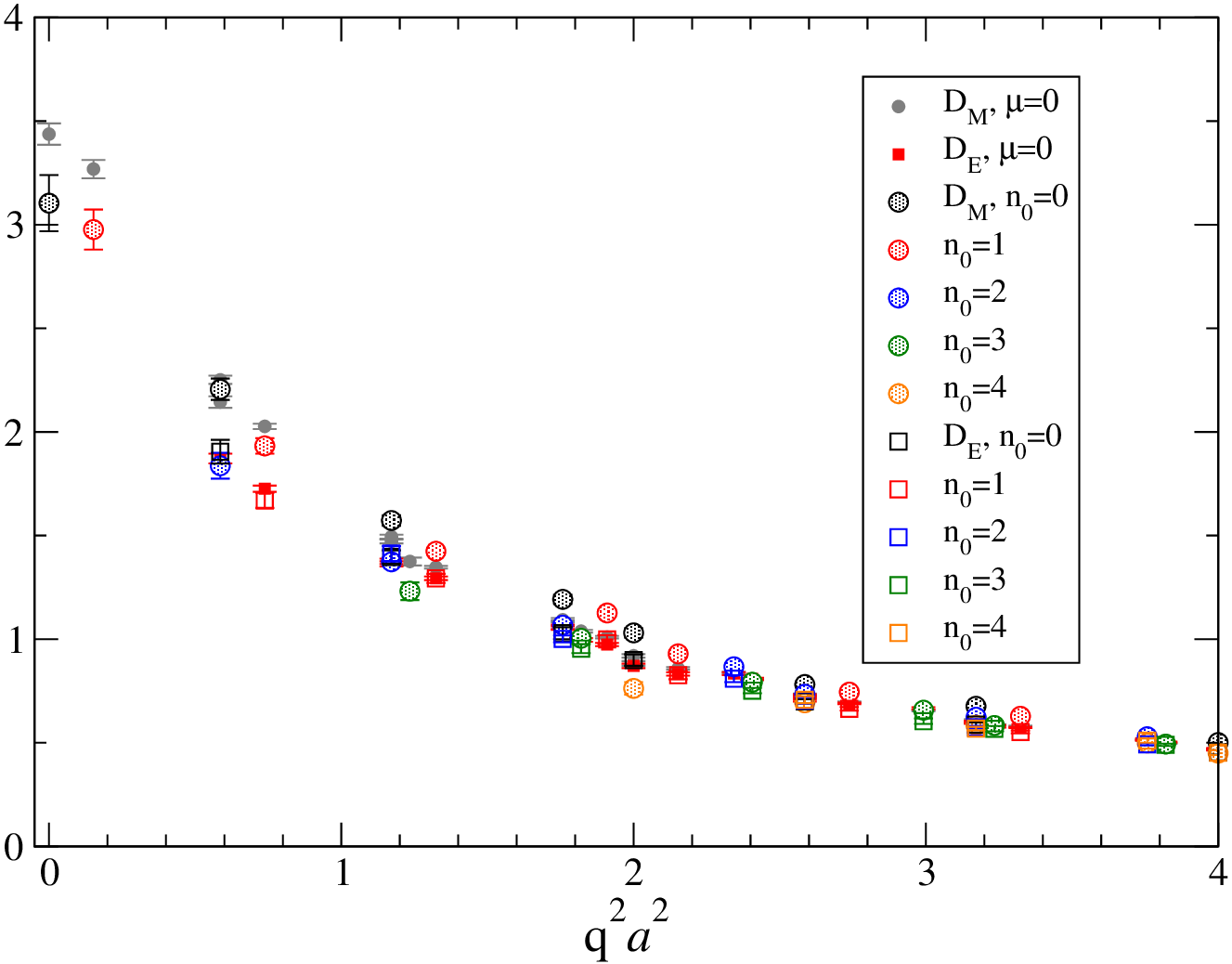}
\caption{Gluon propagator at $\mu=0.7, j=0.04$, together with the
  $\mu=j=0$ propagator for comparison.}
\label{fig:gluon_mu70}
\end{figure}

Figure~\ref{fig:gluon_mu70} shows the gluon propagator at $\mu=0.7,
j=0.04$ as function of the 4-momentum $q^2$, compared to the vacuum
gluon propagator.  We find that the electric propagator remains
virtually unchanged at this point, while some modifications can be
seen in the magnetic propagator.  In particular, the Lorentz or O(4)
symmetry is clearly broken since different values for $q_0$ give
different ``branches''.

\begin{figure}
\includegraphics*[width=\colw]{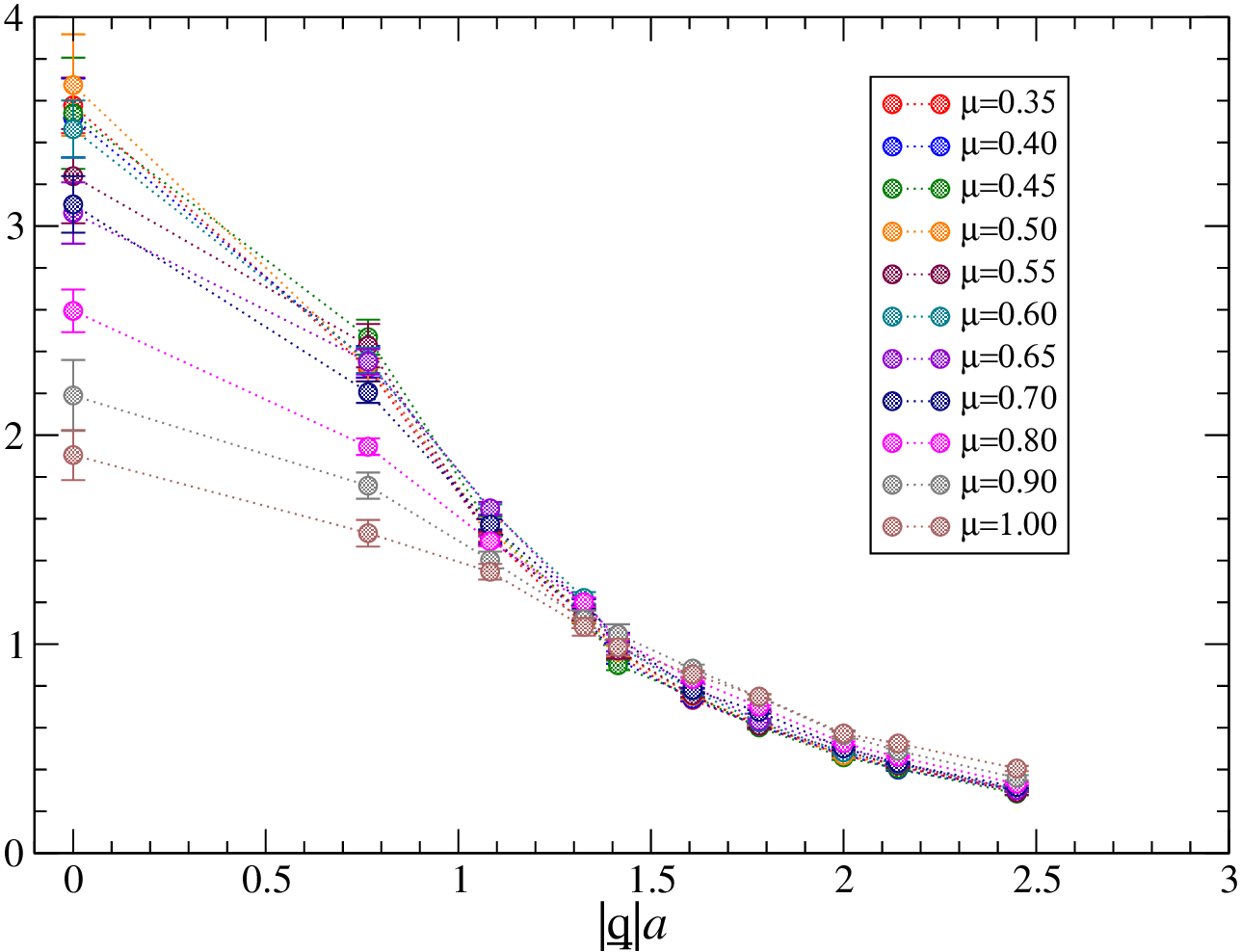}\\
\includegraphics*[width=\colw]{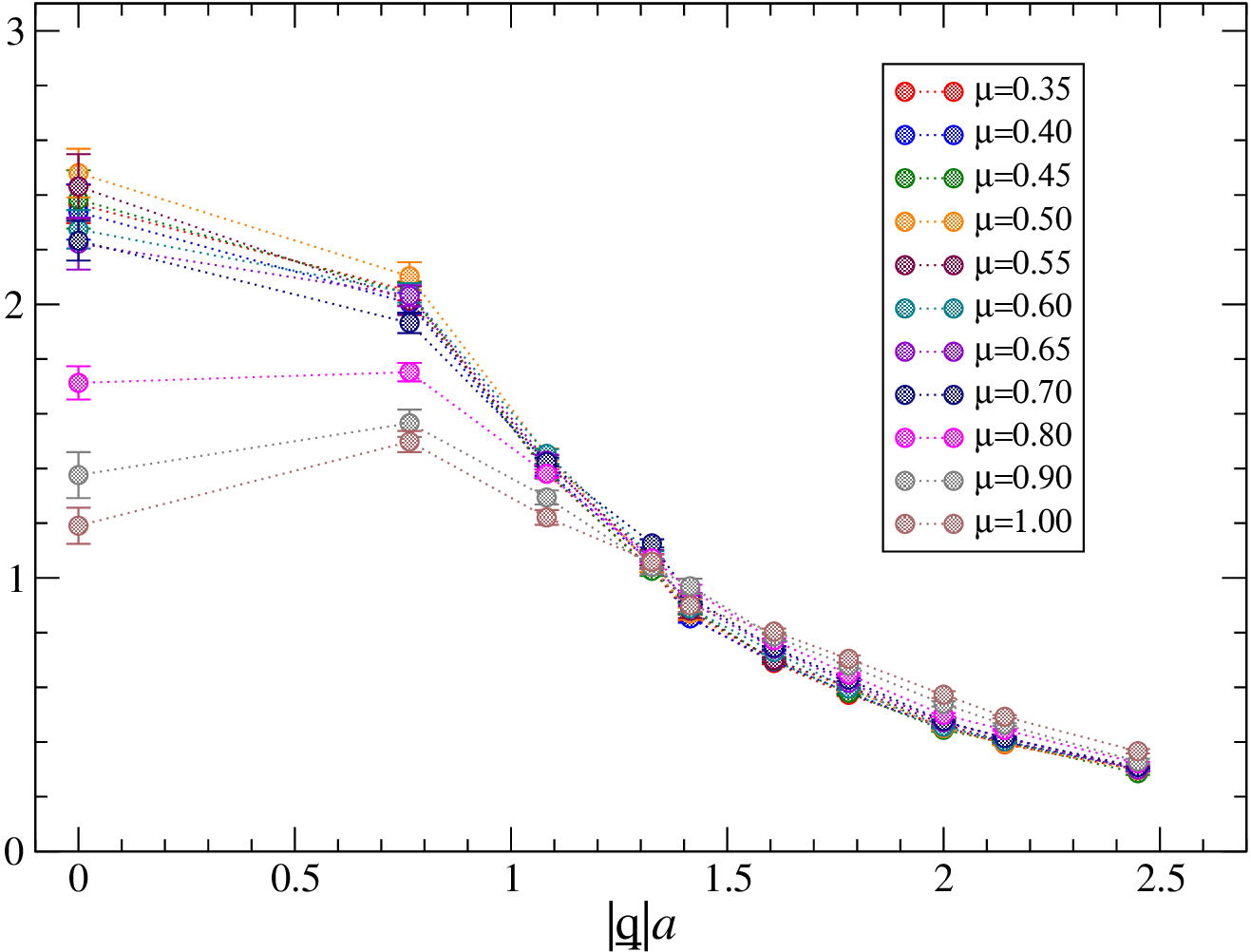}
\caption{Magnetic gluon propagator for $q_0=0$ (top) and $aq_0=2\sin\pi/16$
  (bottom) and diquark source $j=0.04$, as a function of spatial
  momentum $|\vec{q}|$ for various values of the chemical potential.}
\label{fig:DM_k1}
\end{figure}

\begin{figure}
\includegraphics*[width=\colw]{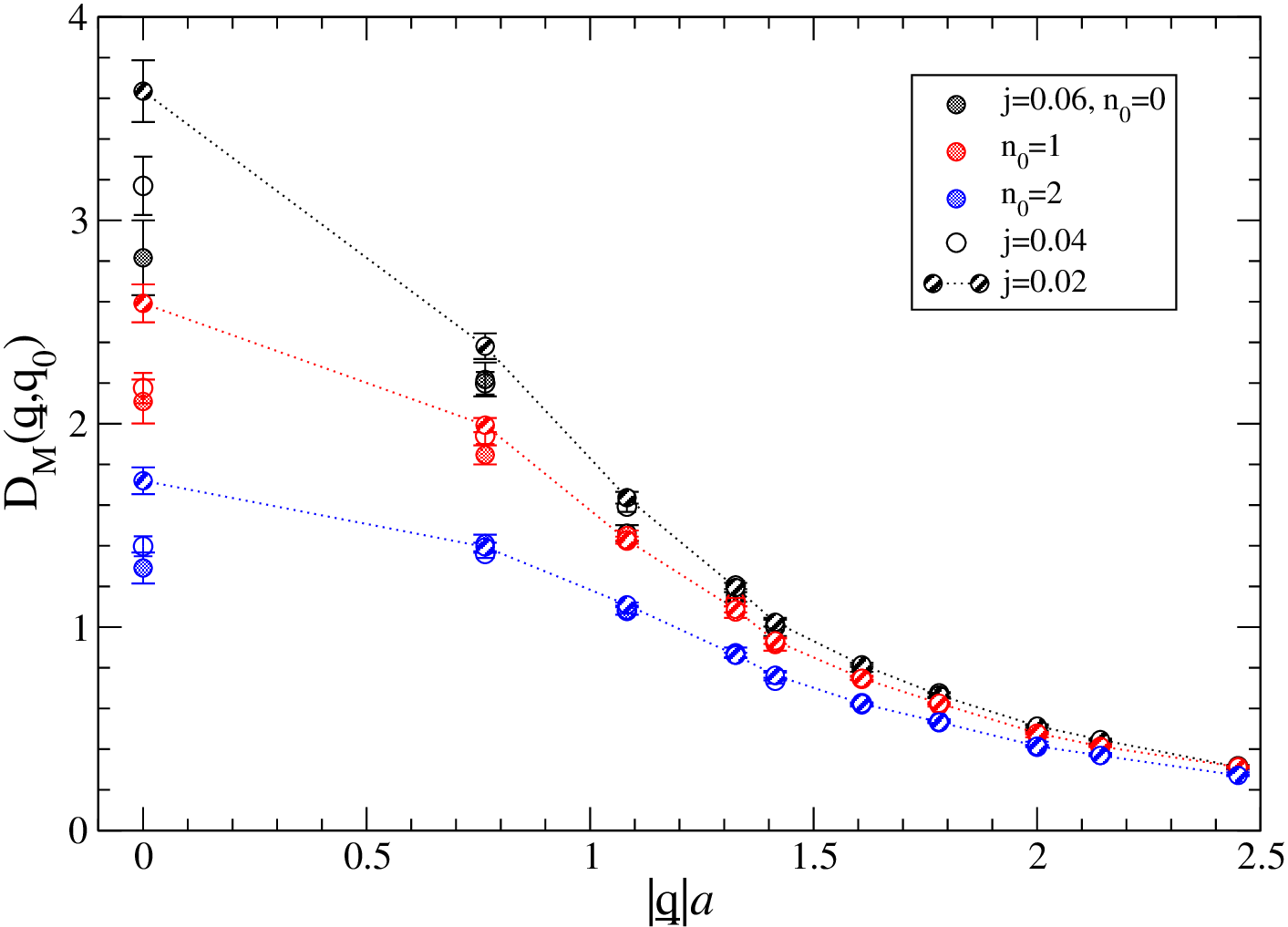}
\caption{Magnetic gluon propagator at $\mu=0.7$ for various values of the
  diquark source $j$.}
\label{fig:gluon_j}
\end{figure}

In order to see the $\mu$-evolution more clearly, we show in
fig.~\ref{fig:DM_k1} the magnetic gluon propagator for the lowest
two Matsubara modes $n_0=0,1$ as function of the spatial momentum
$|\vec{q}|$ for different $\mu$.  We see that there is very little
change up to the deconfinement transition, after which there is a
dramatic infrared suppression and clear ultraviolet enhancement of the
propagator.  These changes makes it look more like an ordinary massive
boson propagator, i.e., the gluon has acquired a magnetic mass
which grows as $\mu$ increases beyond $\mu_d$.
This effect does depend on the diquark source $j$, becoming weaker as
$j\to0$, as figure~\ref{fig:gluon_j} shows.  Simulations at smaller
$j$ in the deconfined phase will be necessary to firm up the picture,
but it seems unlikely that the $j$ dependence seen at $\mu=0.7$ will
be sufficient to wholly cancel the infrared screening effect.  No
dependence on $j$ is seen at $\mu=0.3, 0.5$.

\begin{figure}
\includegraphics*[width=\colw]{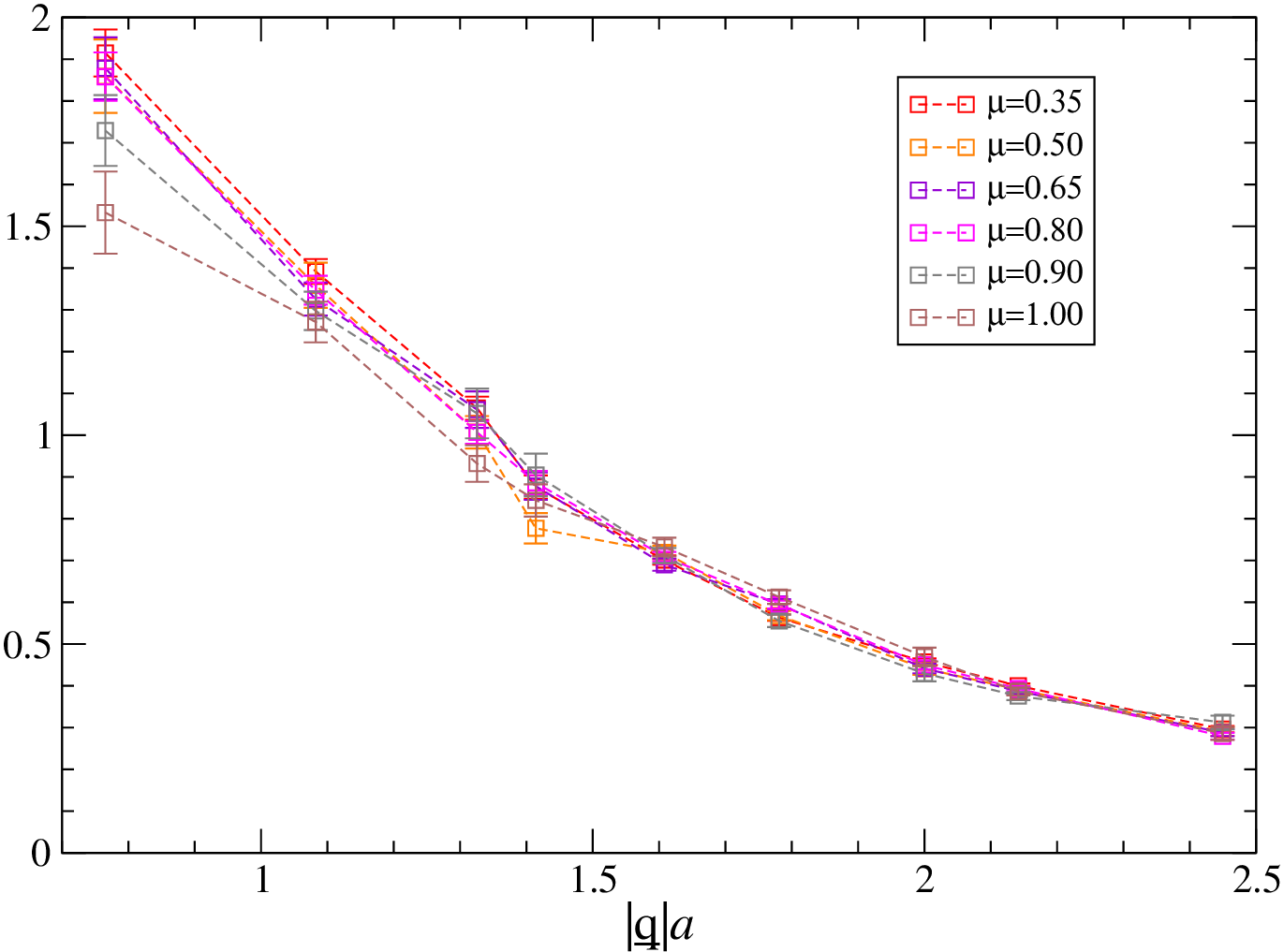}\\
\includegraphics*[width=\colw]{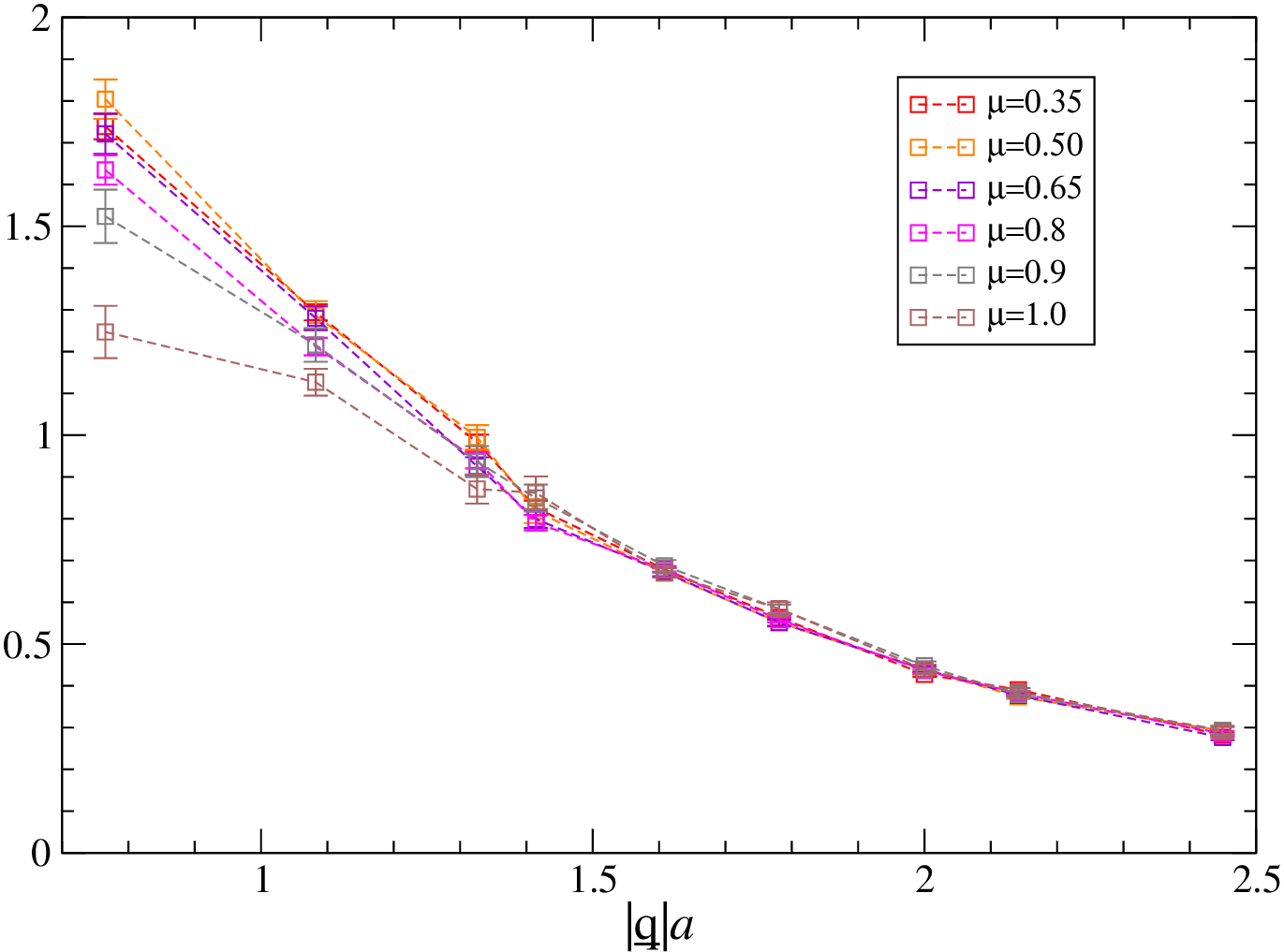}
\caption{As fig.~\ref{fig:DM_k1}, for the electric gluon propagator.}
\label{fig:DE_k1}
\end{figure}

Figure~\ref{fig:DE_k1} shows the electric
propagator for the lowest two Matsubara modes, for selected
values of $\mu$.  While the electric propagator, as seen previously,
remains unchanged up to $\mu=0.7$, it too is clearly suppressed in the
infrared above deconfinement (although to a smaller extent than the magnetic 
case), but appears unchanged in the ultraviolet. No dependence on the
diquark source is seen.  The same effects can be seen in both
static ($q_0=0$) and non-static propagators, although for
$n_0\geq2$ the splitting between electric and magnetic sectors 
becomes harder to detect.  The case of the static
magnetic propagator is particularly interesting, since this is
unscreened to all orders in perturbation theory, yet we observe a
clear screening effect.
The increase of magnetic screening with $j$ also suggests a
non-perturbative origin, hinting at a relation to the non-vanishing
$\qq$ condensate, rather than the presence of light
quasiquark degrees of freedom at the Fermi surface. 
The data in this case may be
distorted by possible finite volume effects, although the consistency
between the $n_0=0$ and $n_0=1$ data indicates that these do not
affect the qualitative picture.

\section{Discussion and outlook}
\label{sec:discussion}

Our results indicate that, at least for the (rather heavy) quark mass
employed in this study, QC$_2$D has three separate phases:
\begin{enumerate}
\item A vacuum phase, for $\mu<\mu_o\approx m_\pi/2$, where the baryon
  density remains zero and all other physical quantities (with the
  likely exception of the hadron spectrum \ckstvz) are unchanged.
\item A confined, bosonic superfluid phase, for $\mu_o<\mu<\mu_d$,
  characterised by Bose--Einstein condensation of scalar diquarks.
  The thermodynamics of this phase can be qualitatively (but not
  quantitatively) described by chiral perturbation theory, and the
  static quark potential is screened by the dense medium.
\item A deconfined phase, for $\mu>\mu_d$, where quarks and gluons are
  the dominant degrees of freedom.  In this phase, a Fermi surface of
  quarks is built up, leading to scaling
  of thermodynamic quantities of the same form as predicted by the
  Stefan--Boltzmann form, but with a different constant of proportionality.  
  We interpret the latter as evidence for a non-zero
  binding energy in this phase, as $k_F>E_F$.  We observe Debye
  screening of the electric gluon propagator, as well as strong
  screening of both static and non-static magnetic gluon modes.
\end{enumerate}
The deconfinement transition occurs at $a\mu_d\approx0.65$, which in
physical units corresponds to $\mu_d\approx600$ MeV. The corresponding quark
density may be estimated as $n_q\approx11\mbox{fm}^{-3}$ directly
from table~\ref{tab:quark}, or from fig.~\ref{fig:eneps} where lattice
artefacts are taken into account, as $\approx$6.5fm$^{-3}$.

Let us discuss some phenomenological implications of our results. First
consider the nature of the deconfinement transition. Since both 
the low density ``nuclear matter'' and high density ``quark matter'' phases are
characterised by a U(1)$_B$ breaking superfluid condensate,
it is tempting to postulate the quark/hadron continuity proposed for 
QCD with $N_f=3$ light flavors in \cite{Schafer:1998ef}. In that case, the
spectrum of physical excitations in both nuclear and quark phases is
qualitatively similar and can be matched using phenomenological insight, for 
instance that the physical state with the smallest energy per baryon is the
6-quark bosonic state known as the H-dibaryon, which can Bose-condense to 
form a superfluid at nuclear densities. In QC$_2$D, however, no matching is
possible. The spectrum in the confined phase consists entirely of bosonic
states, be they $q\bar q$ mesons, $qq$ baryons and their conjugates, or
glueballs. The deconfined phase by contrast has quasiquark excitations with 
half-integer spin and mass gap $\Delta$, the same energy scale
as all non-Goldstone mesonic and baryonic excitations. 
The only states which are conceivably 
lighter are
spin-1 quasigluons, with no counterpart in the low-energy spectrum
predicted by $\chi$PT \ckstvz. Since the spectrum is discontinuous,
it seems natural to suggest that deconfinement in this case is a genuine phase
transition, rather than a crossover.

The most intriguing of our results concern the magnetic gluon
propagator shown in fig.~\ref{fig:DM_k1}. The long-distance screening
observed in the static limit $q_0\to0$ is not predicted at any order
of perturbation theory.  However, this breakdown of perturbation
theory should not be unexpected as the magnetic gluon is an
intrinsically non-per\-tur\-ba\-ti\-ve object.  We have no explanation for the
screening effect, but note that absence of magnetic screening in the
static limit is a crucial ingredient of the celebrated calculation of
the gap scaling $\Delta\propto\exp(-3\pi^2/\sqrt{2}g)$ predicted for
high-density QCD \cite{Son:1998uk}.

\begin{figure}
\includegraphics*[width=\colw]{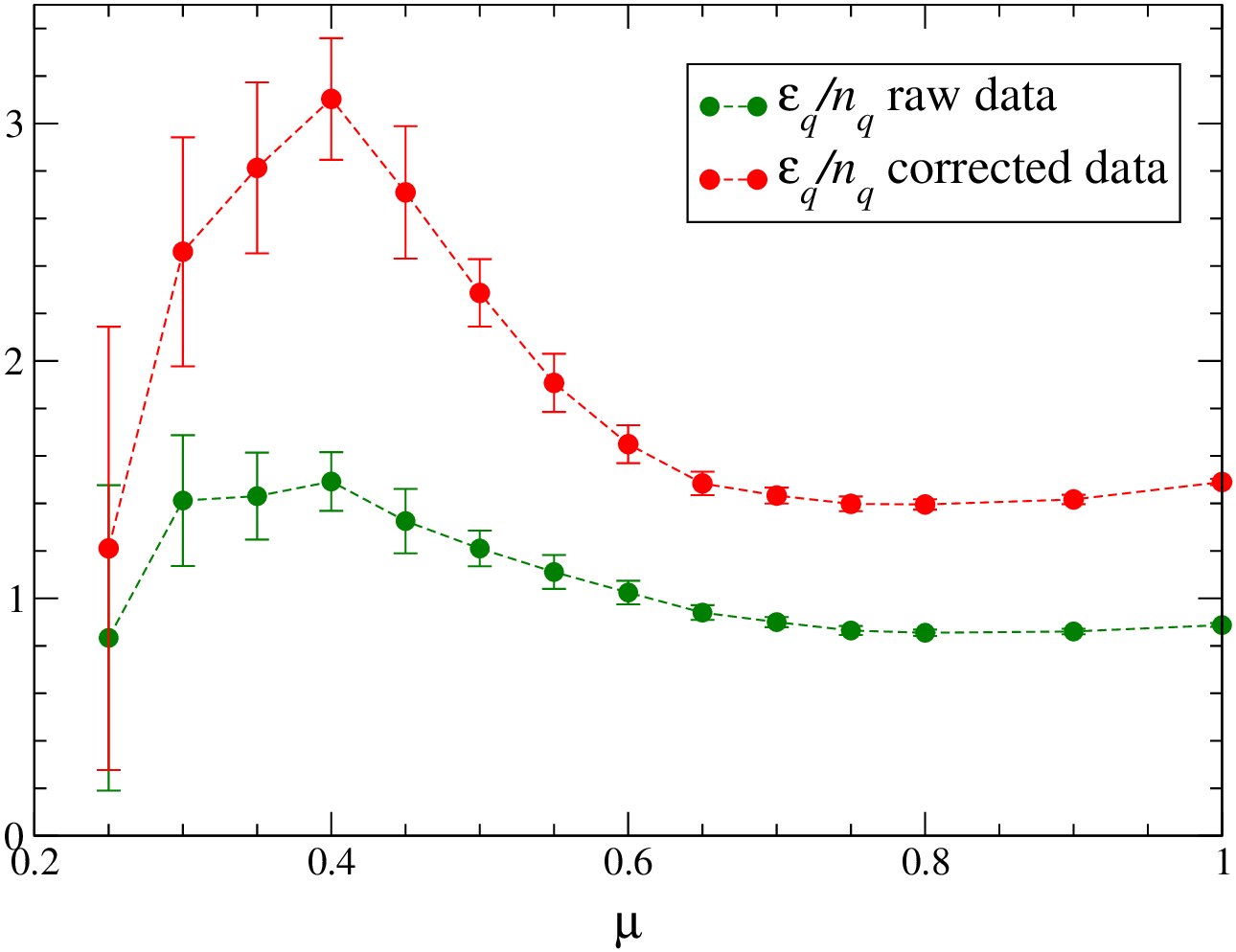}\\
\caption{The ratio $\varepsilon_q/n_q$ vs. $\mu$, using both the raw
data of table~\ref{tab:quark}, and rescaled by a factor
$n_{SB}^\latt/\varepsilon_{SB}^\latt$}
\label{fig:epsonen}
\end{figure}
Finally, it is fascinating to speculate on the astrophysical
consquences of our results. Figure~\ref{fig:epsonen} plots the energy per quark
$\varepsilon_q/n_q$ against $\mu$, and shows a shallow minimum
at $\mu a\approx0.8$. This feature occurs whether or not corrections for lattice
artefacts are applied, and appears to be a robust prediction of our simulation.
Inclusion of the gluon contribution $\varepsilon_g$ shown in
fig.~\ref{fig:energies} will ensure $\partial\varepsilon/\partial n_q>0$ as
$\mu\to\infty$, as of course will recovery of the Stefan--Boltzmann scaling 
(\ref{eq:nSB},\ref{eq:thermoSB}) as asymptotic freedom sets in. Moreover, the
minimum is not predicted by $\chi$PT, 
where $\varepsilon$ increases monotonically
with $n_q$ (see eq.(\ref{eq:DBG})). The minimum resembles a property known as
{\em saturation} in nuclear physics, and implies that objects formed from a
fixed number of baryons, such as stars, will assume their ground states when the
majority of the material lies in its vicinity. Since the minimum lies
above the deconfining transition, we deduce that two-color stars are
made of quark matter.

Orthodox models of quark stars \cite{Glendenning:2000} 
are based on a simple equation of
state such as the Bag Model, which predicts a sharp first order deconfining
transition. The resulting stars have a sharp surface where $p=0$,
along which quark matter coexists with the vacuum. In QC$_2$D by contrast, the 
state of minimum $\varepsilon/n_q$ has $p>0$. A two-color star must therefore
have a thin surface layer, perhaps better described as an atmosphere, formed
from scalar diquark baryons, and whose density falls continuously to zero as the
surface is approached. Matter in the range $0.4\lesssim\mu a\lesssim0.7$, is
mechanically unstable, since according to fig.~\ref{fig:eneps} $\partial
p/\partial\varepsilon<0$ and the resulting sound speed imaginary. At the base of
the diquark atmosphere there will be 
a sharp increase in both pressure and density, and the bulk of the star will be
formed from quark matter with $\mu>\mu_d$. Precise radial profiles, and the
relation between the star mass $M$ and its radius $R$, must await more
quantitative information of the equation of state, which requires the
correct normalisation of $\varepsilon_q$ and $\varepsilon_g$.

We finish by outlining future directions of study.
The lattice spacing used here is quite coarse, and as illustrated in
section~\ref{sec:thermo} lattice artefacts are quite substantial.  It
will be important to repeat the study on finer lattices in order to
gain control over these artefacts.  This is currently underway.

It will also be desirable to find the correct rescaling 
factors for the energy densities, via a non-perturbative determination 
of the Karsch coefficients using simulations on anisotropic lattices \cite{Levkova:2006gn,Morrin:2006tf}.

Since the differences between QCD and QC$_2$D are greatest in the
chiral limit, the heavy quark mass employed here can be considered a
positive rather than a negative.  Nonetheless, it would be desirable
to study a system with lighter quarks, to explore the mass dependence
of our results and interpolate between the r\'egime we have been
exploring and the chiral r\'egime. Better control over the limit $j\to0$ is 
also required.

Beyond the issues considered in this paper, we are intending to study
the hadron (meson and diquark) spectrum and the fate of Goldstone as
well as non-Goldstone modes in the dense medium.  Issues of
interest there are the fate of the vector meson and the possibility
that it becomes light in the dense phase
\cite{Brown:1991kk,Sannino:2001fd,Muroya:2002ry}. 
Also of interest is the pseudoscalar diquark,
which may provide a pointer to the restoration of the U(1)$_A$
symmetry in the medium \cite{Schafer:2002yy}. 
We also intend to study the Gor'kov quark propagator in momentum space,
which will provide information about effective quark masses and the
superfluid diquark gap. Gauge-invariant approaches to 
identifying the presence of a Fermi
surface in a Euclidean simulation can also be employed \cite{Hands:2003dh}.
Finally, once ensembles of the dense medium are
available on a fine lattice, it will be interesting to analyse topological
excitations \cite{Alles:2006ea}, to see to what extent deconfinement in dense 
baryonic
matter resembles deconfinement in a hot medium.

\section*{Acknowledgments}

This work has been supported by the IRCSET Embark Initiative award
SC/03/393Y.  SK thanks PPARC for support during his visit to Swansea
in 2004/05.
We wish to thank Kim Splittorff,
Jimmy Juge, and Mira and Jishnu Dey
for valuable discussions and useful suggestions.
Especially warm thanks go to Shinji Ejiri and Luigi
Scorzato for their participation in the early stages of this project.

\bibliography{lattice,density,gluon}

\end{document}